\documentclass[12pt,a4paper]{article}

\usepackage{amsmath}
\usepackage{amsfonts}
\usepackage{amssymb}
\usepackage{amsthm}

\input{epsf}             
\def\epsfcenter#1{{\vcenter{\hbox{\epsfbox{#1}}}}} 

\def\epsfsize#1#2{0.5#1} 

\newcommand{\sll}{{\rm sl}}   
\newcommand{\SU}{{\rm SU}}
\newcommand{\DSU}{{\rm DSU}}

\newcommand{\SO}{{\rm SO}}

\newcommand{\su}{{\rm su}}
\newcommand{\Uq}{{\rm U}_q}
\newcommand{\dd}{{\mathrm d}}      

\newcommand{\tor}{{\mathrm{tor}}}
\newcommand{\tr}{{\rm tr}}
\newcommand{\Tr}{{\rm Tr}}

\newcommand{\R}{{\mathbb R}}   

\newcommand{\Z}{{\mathbb Z}}   

\newcommand\sixj{6j--symbol}

\newcommand{\curly}[1]{{\mathcal #1}}
\newcommand{\identity}{{I}}  
\newcommand\action{\triangleright}

\theoremstyle{definition}
\newtheorem*{theorem}{Theorem} 
\newtheorem*{lemma}{Lemma} 
\newtheorem*{definition}{Definition}

\begin{document}

\title{The Ponzano-Regge model}

\author{John W. Barrett\\
Ileana Naish-Guzman
\\ \\
School of Mathematical Sciences\\
University of Nottingham\\
University Park\\
Nottingham NG7 2RD, UK\\
}

\date{March 31st, 2009}

\maketitle

\begin{abstract}  
The definition of the Ponzano-Regge state-sum model of three-dimensional quantum gravity with a class of local observables is developed. The main definition of the Ponzano-Regge model in this paper is determined by its reformulation in terms of group variables. The regularisation is defined and a proof is given that the partition function is well-defined only when a certain cohomological criterion is satisfied. In that case, the partition function may be expressed in terms of a topological invariant, the Reidemeister torsion. This proves the independence of the definition on the triangulation of the 3-manifold and on those arbitrary choices made in the regularisation. A further corollary is that when the observable is a knot, the partition function (when it exists) can be written in terms of the Alexander polynomial of the knot. Various examples of observables in $S^3$ are computed explicitly. Alternative regularisations of the Ponzano-Regge model by the simple cutoff procedure and by the limit of the Turaev-Viro model are discussed, giving successes and limitations of these approaches.
\end{abstract}


\section{Introduction}

Three-dimensional quantum gravity can be defined from a number of different points of view. The first of these was the Ponzano-Regge model of quantum gravity on a triangulated 3-dimensional manifold \cite{PR}. This is a state-sum model using the Lie group $\SU(2)$. In a state sum model, there is a quantum amplitude for each assignment of a spin  (irreducible representation of $\SU(2)$) to each edge of the triangulation. The amplitudes are then summed over every possible spin on every edge in the interior of the manifold to give the partition function. The state sum is a discrete version of the functional integral and provides results which are equivalent to the use of functional integral methods.

Since the set of irreducible representations of $\SU(2)$ is infinite, the partition function is often a sum with an infinite number of terms, and in many cases diverges. The main aim of the paper is to discuss the regularisation of this state sum, using various methods, and explaining the extent to which one can succeed. The second part of the paper develops one particular regularisation method and shows how to compute some examples. There is in fact a physical difficulty in that, as we show, the partition function is not always well-defined, but there is a clear cohomological criterion which distinguishes the cases for which the definition succeeds.

A regularisation of this infinite sum using a simple cutoff is proposed in \cite{PR}. This puts an upper limit on the sums over the spin variables and then removes this cutoff with a suitable rescaling. In \cite{PR}, it is shown that this succeeds in a simple example but the
general case was not analysed. In this paper we give an example which remains divergent with this procedure. This means that the resulting limit is not well-defined in every case. It also shows that the simple cutoff regularisation does not lead to invariance under change of triangulation. A variant of the simple cutoff, including an exponential damping as a regulator, is proposed in \cite{KNS}, but no results are known about the limit in which the regulator is removed.

Another regularisation of the Ponzano-Regge model is provided by the Turaev-Viro model, where the Lie group $\SU(2)$ is replaced by its quantum deformation $\Uq\sll2$. When the deformation parameter $q$ is a root of unity, then the regularisation is unnecessary because there are only a finite number of irreducible representations and  the partition function is always well-defined. The regularisation of the Ponzano-Regge model consists of taking a $q\to1$ limit. We show that there is a restricted class of examples where this succeeds and defines a partition function which is independent of the triangulation. However defining and computing a limit $q\to1$ in the general case is not feasible at present.

To define the main regularisation method of this paper we turn to the reformulation of the Ponzano-Regge model in terms of variables in the Lie group $\SU(2)$.
This method of reformulation is well-known for \sixj\ for finite groups \cite{DS} and, more generally, for state sum models for finite-dimensional semisimple Hopf algebras \cite{BWE}.
However the generalisation to $\SU(2)$ has additional subtleties, which are addressed in this paper. The identities relating products of \sixj s to group variables was discovered in \cite{TT} and applied systematically to the Ponzano-Regge model in \cite{BO}. 

This appears to be the appropriate setting in which to regularise the theory. In fact the criterion for the existence of the partition function is expressed in terms of the cohomology classes determined by the flat connections. We do not know how to formulate this criterion directly in terms of the original spin variables, which may explain why regularisation schemes for the spin variables have a limited scope of application.

For a closed manifold, the partition function, when defined, should depend only on the topology of the manifold. For a manifold with boundary, it is usual to fix the variables on the boundary edges, constituting the boundary data of the physical problem, namely specifying a fixed boundary metric. However, the theory is significantly richer if observables are included. In a state sum model this means that the state sum includes a function of the representation variables on some subset of the edges of the triangulation, not necessarily on the boundary. This was introduced first in the Turaev-Viro model \cite{B, BGM}, and then in the Ponzano-Regge model \cite{FL1} by reducing this to the case $q=1$. \footnote{Previously a different set of observables was considered by \cite{T,KS,AW}.} 
 
In this paper a systematic method of defining the Ponzano-Regge partition function with observables is developed. This starts with the reformulation of the partition function in terms of $\SU(2)$ variables. Excess delta functions in the naive, unregulated, formula are removed along a certain choice of a tree of dual edges. This regularisation was proposed for the Ponzano-Regge model in \cite{FLD}. It is also well-known in the theory of Reidemeister torsion, where a standard technique to calculate the torsion for a combinatorial 3-manifold with boundary is to collapse the 3-dimensional dual cells along a tree of dual edges to give a formula for the torsion in terms of the remaining 2-complex (see, for example, \cite{N}). The core of our method, and the correct way to regularise with observables present, is to connect these two methods, constructing the tree by collapsing the 3-manifold from the boundary created by removing cells around the observables. 

The regularisation method has many similarities with a method of fixing a gauge in gauge theories. Ponzano and Regge \cite{PR} \S3 noted that there is an approximate translational symmetry of a \sixj\  which moves one vertex in Euclidean space whilst keeping the remaining vertices of the tetrahedron fixed. Kawamoto, Nielsen and Sato \cite{KNS} found a reformulation of the Ponzano-Regge model by introducing some auxiliary dual variables in which there is an action of the translation group at every vertex of the triangulation (this was re-discovered in \cite{FLD}). However these dual variables are not present in our approach and we do not have the translation group as a gauge group. Thus the regularisation method is not interpreted as a gauge fixing here. In fact the symmetries of 3d quantum gravity are described by the quantum group $\DSU(2)$ \cite{BM, FL2}, and the reconciliation of this with a translation group symmetry is something of a mystery.

The regularised partition function is only well-defined in certain cases, depending on both the topology of the observables and their parameters.
In these cases, the definition reduces to the calculation of a finite-dimensional integral over the space of flat connections, weighted by the Reidemeister torsion of the flat connection and data from the observables. This formula allows us to prove the independence of the definition from both the triangulation of the manifold and from arbitrary choices made in the regularisation procedure, using the well-known topological invariance of the Reidemeister torsion. In many cases the partition function proves to be finite, and a number of examples are calculated explicitly. The case when observables are knotted is examined in detail, following the discovery in \cite{B2} that the observable for torus knots is well-defined only for some values of the parameters of the observables. In these cases, the $q\to1$ limit of the Turaev-Viro torus knot observable was calculated and it agrees with the general formula presented here. A different proof that this partition function, for the case without observables, is independent of the regularisation and the triangulation of a closed manifold appears in \cite{FL2}. One should note that without observables the partition function will only be finite for a limited set of manifold topologies, as discussed in \cite{FL1}.

A different definition of three-dimensional quantum gravity was provided by Witten using his quantization of Chern-Simons quantum field theory \cite{W1, W2}. This quantization corroborates the results here, though in a somewhat indirect way. Firstly, Witten's quantum gravity \emph{with} a cosmological constant is known to be equivalent to the Turaev-Viro model, since both are the square of an $\SU(2)$ Chern-Simons theory \cite{W1,R}. This suggests, very heuristically, that Witten's quantum gravity \emph {without} a cosmological constant should give the Ponzano-Regge model. 
In fact, Witten's gravity without cosmological constant gives a partition function (without observables) on a 3-manifold which is defined in terms of the analytic torsion of the space of flat connections. The Cheeger-M\"uller theorem asserts that the analytic torsion is equal to the Reidemeister torsion \cite{BV}, so the partition function should agree with our formula. Some further comments on the relation between the two approaches are made in section \ref{conclusion}.

The paper starts in section \ref{PRSS} with the definition of the Ponzano-Regge model. There is a short section on the use of the Turaev-Viro model to regularise it, though the remainder of the paper is independent of the Turaev-Viro model. Then the Ponzano-Regge model is reformulated in terms of integrals over group elements located on triangles (or dual edges). 

The main definition of the partition function with observables 
is then given in section \ref{group}. These observables are specified
by giving a conjugacy class in $\SU(2)$ to each edge of a graph embedded in the manifold.
The partition function is then defined by restricting the holonomy of the group elements around such an edge to lie in the given conjugacy class. 
 It is shown in section \ref{EPF} that the criterion for the formula to make sense is that the second twisted cohomology group should vanish at each point of the integration over the group elements. 

If this criterion is satisfied, then in section \ref{invariance}, the resulting expression is written in terms of the Reidemeister torsion, which shows that the partition function is independent of both the regularisation which is used in the definition and the triangulation of the manifold. The relation between the Ponzano-Regge partition function and Reidemeister torsion was previously announced in \cite{BNG}. A proof of the independence from regularisation and triangulation for the case without observables was given previously, using different methods, in \cite{FL2}.

Some examples for observables in $S^3$ are calculated in section \ref{calculation}, and particular features of both knots and planar graphs are explored. The relation of the partition function for a knot to the Alexander polynomial of the knot, discovered in \cite{B2}, is proved in general.

\section{The Ponzano-Regge state sum}\label{PRSS}

\subsection{The weight}\label{theweight}
\begin{figure}[hbt]
$$\epsfbox{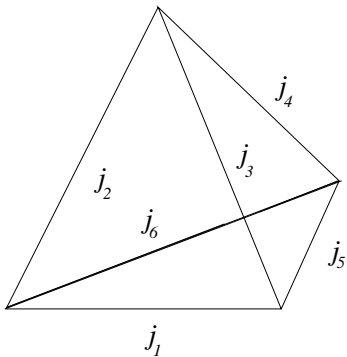}$$
\caption{The tetrahedron}\label{tet}\end{figure}

Let $M$ be a triangulated compact 3-manifold. A state of the model is an assignment of an irreducible representation of $\SU(2)$ to each edge of the triangulation. The irreducibles are labelled by a non-negative half-integer parameter $j$, called the spin, so that the dimension of the representation is given by $2j+1$. So one can think of a state as an assignment of these half-integers to the edges. For each state there is a certain \emph{weight}, a real number. The weight is given by the local formula
\begin{equation}\label{weight}
W=\prod_{\text{interior edges}}(-1)^{2j}(2j+1) \prod_{
\text{interior triangles}}(-1)^{j_1+j_2+j_3}\prod_{\text{tetrahedra}}\left\{\begin{matrix}j_1&j_2&j_3\\j_4&j_5&j_6\end{matrix}\right\}.
\end{equation}
In this formula, the weights for each simplex are determined by the labels on that simplex. The edges and triangles which appear are those that are in the interior of the manifold, i.e., not on the boundary. The weight for a tetrahedron is a $6j$-symbol, for which the formula is given in \cite{PR}. 
To use the $6j$-symbol  correctly, the spin labels are those labelling the edges of the tetrahedron according to figure \ref{tet}.
The $6j$-symbol is symmetrical under the group of permutations of the vertices of the tetrahedron, so it does not matter in which of the 24 possible ways this is done. 

The weight is defined to be equal to $0$ if the \emph{admissibility conditions} are not satisfied. The admissibility conditions are that for each triangle 
\begin{equation} \label{admeven} j_1+j_2+j_3\in\Z,\end{equation}
\begin{equation} \label{admtriangle} j_1+j_2-j_3\ge0,\end{equation}
and two other similar equations obtained by permuting 1,2 and 3. In fact it is usual to define the value of a $6j$ symbol to be zero if any of these conditions are not satisfied on its faces.

The signs associated to the triangles can be absorbed into the tetrahedron weight, by defining
$$\left\|\begin{matrix}j_1&j_2&j_3\\j_4&j_5&j_6\end{matrix}\right\|=
i^{2(j_1+j_2+j_3+j_4+j_5+j_6)}
\left\{\begin{matrix}j_1&j_2&j_3\\j_4&j_5&j_6\end{matrix}\right\}.$$
This leads to an alternative formula for the weight
\begin{equation}\label{weightalt}
W'=\prod_{\text{interior edges}}(-1)^{2j}(2j+1) \prod_{\text{tetrahedra}}\left\|\begin{matrix}j_1&j_2&j_3\\j_4&j_5&j_6\end{matrix}\right\|.
\end{equation}
which agrees with (\ref{weight}) up to a power of $i$ which depends only on the boundary data. The signs associated to the edges cannot be removed in any similar way.\footnote{Many recent papers have omitted the sign factors. Such definitions lead either to incorrect formulae or, at best, give a state sum which depends on the triangulation.}

Ponzano and Regge's original paper only actually gave a formula for the sign factor in the state sum for the case where $M$ is a three-dimensional ball. In that case, their formula agrees with either $W$ or $W'$ up to a power of $i$ which depends only on the boundary data.

\subsection{The partition function}
The partition function, or state sum, is obtained by summing over all values of the variables, subject to fixed values on the boundary.
\begin{equation}\label{PR} Z=\sum_{j_1, j_2, \ldots j_n} W,\end{equation}
where $j_1$, $j_2$, etc., are the spins associated to the edges in the interior of the manifold. For some triangulations this gives a finite sum, whereas for some other triangulations this is a divergent infinite sum. It is possible that there are triangulations where the sum is infinite yet convergent, but we do not know of any examples. Therefore in general a regularisation is required to make (\ref{PR}) finite. Two possible regularisation methods are discussed in the rest of section \ref{PRSS}, and the main method of this paper is presented in section \ref{group}. The main requirement of a good definition of the partition function is that it should be independent of the triangulation of the interior of the manifold.

The physical interpretation of the partition function is obtained by regarding a spin label $j$ on an edge as specifying a metric length, $j+1/2$. Then the admissibility conditions imply that each triangle has a non-degenerate Euclidean geometry. The fixed data on the boundary amounts to a 2-dimensional Euclidean metric on this boundary, flat on the interior of each triangle. The partition function is therefore a sum over metrics on the interior of the manifold which agree with the fixed boundary metric. 

A tetrahedron also has a flat geometry determined by its edge lengths, but this can be either Euclidean or Lorentzian signature. According to the Ponzano-Regge asymptotic formula, the weight of each 3-metric is related to the Einstein action without cosmological constant. If a tetrahedron is Euclidean the weight is oscillatory. If it is Lorentzian then it is suppressed by a real exponential factor with the Lorentzian action \cite{BF}. The latter case is interpreted as a tunnelling phenomenon.

\begin{figure}
$$\epsfbox{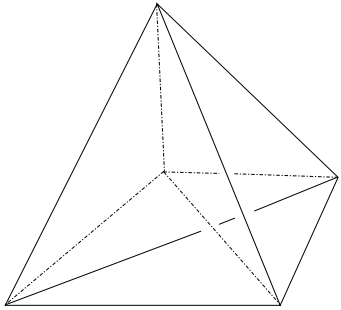}$$
\caption{Subdivided tetrahedron}\label{fourtet}
\end{figure}

For many triangulations the partition function diverges, and so a regularisation is required. An example is given by the computation of the partition function for the triangulation of the three-ball obtained by dividing a tetrahedron into four by adding an extra vertex in the centre (figure \ref{fourtet}).\cite{PR}
The boundary of this triangulation is identical to the boundary of a tetrahedron.
This led Ponzano and Regge to suggest the `simple cutoff' regularisation procedure by putting a cutoff on the maximum spin in the sum, so that $0\le j\le\Lambda$, rescaling by a certain $\Lambda$-dependent factor $M_\Lambda$ for each internal vertex, and then taking the limit 
$$Z=\lim_{\Lambda\to\infty} M_\Lambda^{-v}\sum_{j_1, j_2, \ldots j_n=0}^\Lambda W,$$
with $v$ the number of interior vertices.
For this example, the regularisation converges and gives a partition function which is exactly equal to the partition function for the tetrahedron, as required by triangulation independence. 
An example in which the limit does not converge, however, is discussed in section \ref{Bing}.

\subsubsection{Pachner moves}

Any two triangulations of a 3-manifold can be related by a sequences of local moves on the triangulation known as Pachner moves. Therefore a strategy for proving triangulation independence is to show that the state-sum model is invariant under the Pachner moves. In three dimensions, there are just two Pachner moves which, together with their inverses, generate all triangulations. The 4-1 move replaces the subdivided tetrahedron of figure \ref{fourtet} with a single tetrahedron. The other move is known as the 3-2 move and replaces three tetrahedra with a common interior edge with two tetrahedra glued together on a common triangle. For details of these moves and their use in state sum models of Turaev-Viro type, see 
\cite{BWPL}.

In the context of the Ponzano-Regge model, the 3-2 move is true for the unregularised partition function (\ref{PR}), the state-sums on both sides being finite, and is known as the Biedenharn-Elliot identity. The 4-1 move is not true without regularisation, as just discussed. As demonstrated by Ponzano and Regge, the 4-1 move is exactly true with their simple cutoff regularisation. The problem is that the Biedenharn-Elliot identity is no longer necessarily true with the cutoff in place, and so both moves cannot always be applied. For this reason the discussion of invariance in this paper is not based on the Pachner moves applied to \sixj s.

\subsubsection{The dual complex and spin foams} The partition function weight $W$ can be formulated equivalently using the 2-skeleton of the dual of the triangulation. This leads to a generalisation in which the state sum can be formulated on more general 2-complexes.

For a triangulation of a closed manifold, the description of its dual complex is quite straightforward. The dual complex has a $k$-cell \footnote{A $k$-cell is topologically a $k$-dimensional ball, an example of which is $[0,1]^k$. The boundary of the ball is included.} for each $3-k$-simplex of the triangulation. There is a vertex of the dual complex at the barycentre of each tetrahedron, and the dual 1-cell corresponding to a given triangle connects the two vertices dual to the two tetrahedra having that triangle as a face. It meets the triangle at one point. This pattern continues up the dimensions, with a dual 2-cell attached to the 1-skeleton (the 1- and 0-cells) and intersecting an edge of the triangulation at one point. Dual 3-cells fill in around each vertex of the triangulation, so that the dual complex is a decomposition of the manifold. The dual complex can be constructed in a precise way by first forming the barycentric subdivision of the triangulation and then grouping simplexes of this subdivision together to form the dual cells (as in the transition from figure \ref{rn-triangulation} to figure \ref{rn-dual}). The term $n$-skeleton refers to the subcomplex formed by the set of cells up to and including dimension $n$.

\begin{figure}[hbt]
$$ \epsfbox{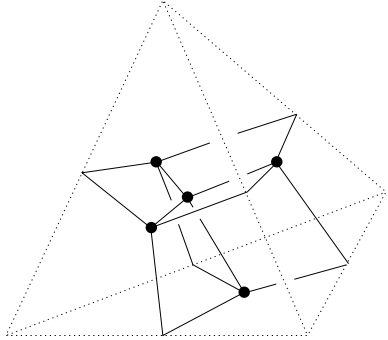} $$
\caption{The dual of a tetrahedron}\label{dualtet}
\end{figure}

For a triangulation of a manifold $M$ with boundary, the dual complex has cells which are dual to the simplexes of $M$, and, in addition, cells which are dual in $\partial M$ to the simplexes in $\partial M$. This is illustrated by the following simple example of the dual complex for the tetrahedron (figure \ref{dualtet}). The number of 0-,1-,2- and 3-cells in this complex is 5,10,10,4, and of these, 4,6,4,0 are on the boundary.
The spin variables in the weight (\ref{weight}) can be associated to the dual 2-cells, and the weight factors in $W$ correspond to the interior dual 0- 1- and 2-cells, which taken together, form the dual 2-skeleton. It can be seen that the dual 3-cells are not required to formulate $W$, but they are in fact uniquely determined by the dual 2-skeleton. The dual 2-skeleton with the spin variables on its 2-cells is called a spin foam. 

This concept of a spin foam can be generalised to a 2-complex which is {\em locally} like the dual of a triangulation. In other words, each point has a neighbourhood isomorphic to the neighbourhood of a point in figure \ref{dualtet}. Such 2-complexes are called special polyhedra \cite{cxbook}. In this case the complex does not have to be dual to a triangulation. This formulation of a spin foam is more flexible than triangulations; it includes, for example, the duals of degenerate triangulations, and is often useful to give simpler expressions for partition functions.

The boundary edges of a spin foam form a trivalent graph with each edge inheriting a spin variable from the interior 2-cell which meets it. This graph lies in the boundary surface, and is called a generalised spin network. The term spin network itself is reserved for the slightly different concept of a graph in three-dimensional space and its projection onto the plane. This is introduced below, in section \ref{spinnetworks}.

\subsection{Limit of Turaev-Viro}
A more sophisticated regularisation is provided by the Turaev-Viro model \cite{TV}. This section shows how this regularisation can be used to prove a result on the independence of the Ponzano-Regge state sum from the choice of triangulation. The rest of the paper is however independent of the Turaev-Viro model, and the reader can skip to section \ref{spinnetworks}.

In the Turaev-Viro model there is an additional parameter $q=e^{i\pi/r}$ for integer $r\ge3$, and the weight $W_q$ is given by a formula analogous  to (\ref{weightalt}) in which every factor depends on $q$. Crucially, the spins are limited to the range $0\le j\le (r-2)/2$, and there is an extra admissibility condition
$$j_1+j_2+j_3\le r-2.$$ 

The $6j$-symbol is replaced by a quantum $6j$-symbol, and the factor for each edge by a quantum dimension. As $r\to\infty$, $q\to1$ and 
$$W_q\to W'.$$
This follows from the fact, which is easy to check from the formulae in \cite{TV}, that the factor for each simplex converges to that in  (\ref{weightalt}). As a result, the formula $(-1)^{2j}(2j+1)$ is called the quantum dimension of the corresponding representation of $\SU(2)$.

The Turaev-Viro model is defined by the partition function
 $$Z_q=N_q^{-v}\sum_{j_1, j_2, \ldots j_n=0}^{(r-2)/2} W_q,$$
where $N_q$ is a constant depending on $q$, and $v$ is the number of internal vertices. Since the sum is finite, this is always well-defined. Moreover, the partition function is independent of the triangulation of the interior of $M$; $Z_q$ depends only on the boundary triangulation, the boundary data, and the topology of $M$.

The Turaev-Viro regularisation is superficially similar to the regularisation proposed by Ponzano and Regge \cite{OS}. It seems clear that the upper limit to the sums, $(r-2)/2$, plays the role of the Ponzano-Regge cutoff, and the constant $N_q$ plays the role of the Ponzano-Regge scaling factor for each internal vertex. However it is not straightforward to take the limit $q\to1$ of $Z_q$. This is due to the fact that for many triangulations the number of terms in the sum increases without limit as $q\to1$.

\subsubsection{Non-tardis triangulations}

One case in which the limit can be taken is for a certain set of triangulated manifolds with boundary. These are the triangulations for which the Ponzano-Regge state sum is constrained to be a finite sum by the boundary data and the admissibility conditions. This is in fact a purely combinatorial condition, i.e. it depends only on the triangulation and not on the boundary data. 

Given a triangulation of a manifold with boundary, the tardis\footnote{The TARDIS is a box in a science-fiction series of stories which is much bigger on the inside than its external geometry would suggest.} is defined to be the set of edges for which the state sum is not constrained to be a finite sum by the admissibility conditions with fixed boundary data. A non-tardis triangulation is one for which the tardis is empty.

For a non-tardis triangulation, the state sum is, by definition, a finite sum. For these, the Ponzano-Regge state sum is defined directly by (\ref{PR}), and this formula can be compared to the limit of the Turaev-Viro partition function as $q\to1$.

The tardis of a triangulation can be recognised as follows. The triangle inequalities (\ref{admtriangle}) imply that one edge of a triangle has bounded spin if the other two edges do. Therefore one can proceed by iteration as follows. Define a sequence of subsets of the set of all edges, 
starting with $E_0$ being the set of boundary edges. If there is an edge which is not in $E_k$ but lies on a triangle with its two other edges in $E_k$, then add this edge to $E_k$ to form $E_{k+1}$. When this terminates at $E=E_n$, the remaining edges (i.e., those not in $E$) form the tardis. An example of this is when there are interior vertices, as in figure 2. The edges meeting the interior vertices belong to the tardis. Therefore a non-tardis triangulation always has no interior vertices.

This can be formulated alternatively in terms of the 2-skeleton of the dual cell decomposition to the triangulation. 
Start with $C_0$ being the 2-complex generated by the 2-cells which are dual to the interior edges of the triangulation (i.e. these 2-cells together with the 0- and 1-cells in their boundary). Then the transition $C_k\to C_{k+1}$ is a collapsing move (see section \ref{collapsing}) in which one 1-cell $\sigma$ and one 2-cell $\tau$ are removed. The 1-cell $\sigma$ lies in the boundary of $\tau$ and is not in the boundary of any other 2-cell of $C_k$. Thus it is on the `boundary' of the 2-complex.  At each stage the 2-cells in $C_k$ are dual to the set of edges of the triangulation not in $E_k$.

The collapsing continues until no more collapses are possible; the remaining 2-cells of the final $C_n$ form a 2-complex $C$ with no boundary.   These 2-cells, dual to the edges of $E$, will also be referred to as the tardis of the triangulation. According to the `helter skelter' argument of Lickorish, collapsing moves carried out in any order will always lead to the same non-collapsible final result, so that the tardis is uniquely determined. 

In particular, the dual 2-skeleton collapses to a 1-complex if and only if the triangulation is non-tardis. For these triangulated manifolds, it is shown in the proof below that the limit of the Turaev-Viro model is equal to the Ponzano-Regge partition function. This can be used to prove a result about the triangulation-independence of the Ponzano-Regge model.

\begin{theorem}
Let $M$ be a manifold with a given triangulation of its boundary and fixed spin labels on these boundary edges. Then any two non-tardis triangulations which extend the boundary triangulation to the interior of $M$ determine the same Ponzano-Regge partition function. \end{theorem}

\noindent\emph{Proof.}
For a non-tardis triangulation, in the limit $q\to1$, 
eventually the Turaev-Viro state sum contains the same finite number of terms as the Ponzano-Regge sum, and for each term $W_q\to W'$. Also, $v=0$. Therefore $Z_q\to Z'=\sum W'$. Since the Turaev-Viro partition function is independent of the interior triangulation, it follows that the Ponzano-Regge partition function is also.

Note that this result only applies to a given manifold with triangulated boundary, if there exists a non-tardis triangulation. However where one does exist, it provides a canonical choice of definition of the partition function. In the case of a 3-ball, with boundary $S^2$, the result was previously known, because the Ponzano-Regge state sum gives just the spin network evaluation on the boundary \cite{PR}. This is described explicitly by the algorithm of Moussouris \cite{M}.

The Biedenharn-Elliot identity (3-2 Pachner move) can be applied directly to the Ponzano-Regge partition function in the case of a non-tardis triangulation. This would prove some limited number of cases of triangulation independence. Any such results are however included in the theorem above. 

\subsection{Bing's house}\label{Bing}
Bing's house is an interesting example of a spin foam which exhibits some surprising and perhaps pathological features of the Ponzano-Regge model.

\begin{figure}[htb]
{\def\epsfsize#1#2{0.25#1}$$\epsfbox{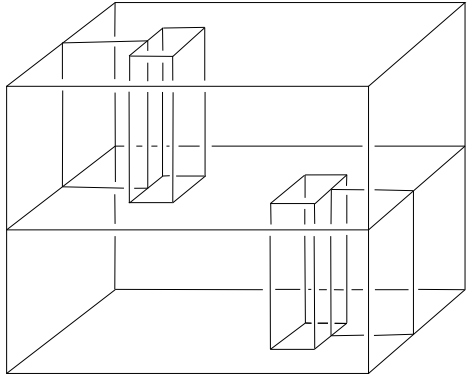}$$}
\caption{Bing's house. The lower room is accessed from the outside via the chimney in the roof, while the upper room is accessed similarly from underneath the house.}
\label{binghouse}
\end{figure}

Not all triangulations of the 3-ball with no interior vertices are non-tardis triangulations. For example, one can construct triangulations so that the interior edges are dual to a non-collapsible 2-complex. An example of a non-collapsible 2-complex is given by Bing's house with two rooms \cite{RS}, depicted in figure \ref{binghouse}. 
\begin{figure}[htb]\label{bingskeleton}{\def\epsfsize#1#2{0.25#1}$$\epsfbox{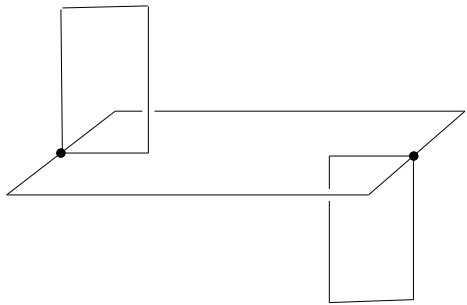}$$}
\caption{The 1-skeleton of Bing's house.}
\end{figure}
Triangulations with tardis homeomorphic to Bing's house can be constructed by thickening the 2-cells (`walls') of Bing's house to a three-dimensional manifold $M_B$, which is easily seen to be the 3-ball. This thickening is constructed by centering a tetrahedron on each of the two vertices of Bing's house (see figure \ref{bingskeleton}) and then dividing the thickened wall panels into as many tetrahedra as necessary, without introducing any interior vertices, to connect up the two tetrahedra. If the spin labels on the boundary of this spin foam are all 0, then the partition function reduces to the Ponzano-Regge model for the Bing's house 2-complex itself. The formula is
$$ Z=\sum_{c=0}^\infty\sum_{b=0}^{2c}\sum_{a=0}^{2c} (-1)^{a+b}(2a+1)(2b+1)(2c+1)\left\{\begin{matrix}a&c&c\\c&c&c\end{matrix}\right\}\left\{\begin{matrix}b&c&c\\c&c&c\end{matrix}\right\}.$$
As there are no interior vertices, then the Ponzano-Regge simple cutoff regularisation has no rescaling by powers of $M_\Lambda$. Nevertheless the sum is infinite and it is not absolutely convergent, as follows from an application of the Ponzano-Regge asymptotic formula \cite{R2}. Scaling all of the spins in the summand by a factor $\lambda$ gives a scaling of $\lambda^{-3/2}$ for each \sixj, and so the absolute value of the summand is $O(1)$ in this limit.
 The fact that the sum is not absolutely convergent means that it is possible to rearrange the terms in the sum to get different answers. Indeed, taking the sums over $a$ and $b$ first, and using the identity
\begin{equation}\label{6jidentity}\sum_{a=0}^{2c}(-1)^{a}(2a+1)\left\{\begin{matrix}a&c&c\\c&c&c\end{matrix}\right\}
=\delta_{c0},\end{equation}
results in $Z=1$. However rearranging the sum by using the Ponzano-Regge cutoff gives, according to a numerical calculation\footnote{This calculation was first done for us by Stefan Davids.}, a divergent limit, as shown in figure \ref{divergence}. The figure shows the cutoff partition function $Z_\Lambda$ against the cutoff $\Lambda$, where $Z=\lim_{\Lambda\to\infty}Z_\Lambda$.
\begin{figure}[htb]
\centerline{
\vbox{\hbox{$Z_\Lambda$}\kern.1in\hbox{\epsfbox{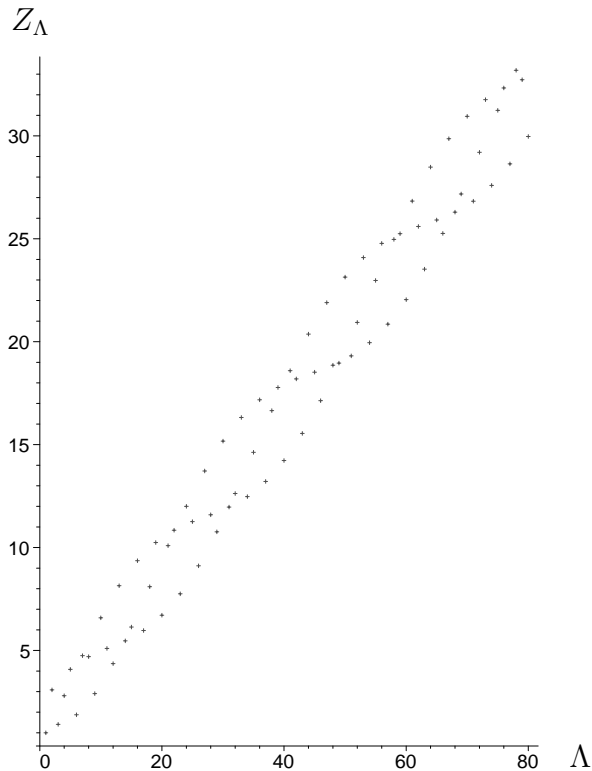}\quad$\Lambda$}}
} 
\caption{The partition function of Bing's house with simple cutoff.}\label{divergence}
\end{figure}
The divergence is essentially due to the fact that the cutoff truncates some of the sums over $a$ and $b$ before the upper bound $2c$ is reached.

By contrast, the Turaev-Viro partition function for this spin foam is equal to 1 for all $q$, and so the Turaev-Viro regularisation for the Ponzano-Regge model is also 1. This example shows that the Turaev-Viro regularisation is better behaved than the simple cutoff regularisation. 

Another corollary of this example is that the simple-cutoff regularisation of the Ponzano-Regge model is not independent of the triangulation. For example, let $M_B$ be the triangulation of the 3-ball constructed above. One can construct non-tardis triangulations of the 3-ball with the same boundary (e.g. using the algorithm of Moussouris). Then the partition function for this triangulation is a finite sum, as discussed in the previous section, whereas it is not convergent for $M_B$.

\subsection{Spin networks}\label{spinnetworks}
The Ponzano-Regge weights can be understood more systematically from the point of view of spin networks.
A spin network is a diagram which encodes a calculation using intertwining operators for representations of $\SU(2)$. In these calculations, the order of various arguments is important and the diagram is a good way of keeping track of these. In fact the key to the application of representation theory to physics starts with the observation that many of the identities satisfied by spin networks correspond to deforming these diagrams. The diagrammatic approach was pioneered by Penrose \cite{P}, and then extended in the context of the $q$-deformation by Kauffman \cite{K}. The operator representation of spin network diagrams is also reviewed in \cite{MA}. A brief outline is explained here, concentrating on the most puzzling features. 

If $a$ and $b$ are representations of $\SU(2)$, then an intertwining operator from $a$ to $b$ is represented by a diagram drawn in a rectangular subset of the plane, with boundary given by two horizontal edges and two vertical edges. This rectangular subset has its bottom edge labelled by $a$ and its top edge labelled by $b$. 

\begin{figure}
$$\epsfbox{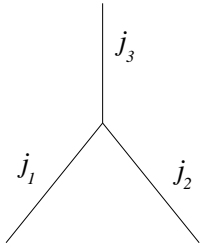}$$
\caption{The spin network vertex}\label{spinnetvertex}
\end{figure}

Inside the diagram there are a number of lines labelled by irreducible representations and these lines are allowed to meet the boundary of the rectangular region at the bottom edge or the top edge. For example, if the bottom edge of the diagram meets two lines labelled with $j_1$ and $j_2$ (as in figure \ref{spinnetvertex}), then the bottom label of the diagram is $a=j_1\otimes j_2$. Two diagrams are regarded as the same if one is obtained from the other by a continuous deformation of the plane which keeps the horizontal and vertical edges intact.

A triple of spins $(j_1,j_2,j_3)$ is called admissible if representation $j_3$ occurs in the Clebsch-Gordan decomposition of $j_1\otimes j_2$ (see section \ref{theweight}).
For each admissible triple $(j_1,j_2,j_3)$, there is a canonical choice of intertwining operator $j_1\otimes j_2\to j_3$. This is a basis vector in the one-dimensional space of such intertwiners, and is represented as the diagram in figure \ref{spinnetvertex}.

The intertwiners are composed by stacking the rectangles vertically, and tensored together by stacking the rectangles horizontally. 
When composing the diagrams vertically, the lines have to meet to form continuous line segments through the join. For example, composing the intertwiner
$$\epsfbox{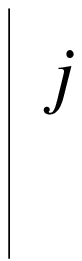}\quad\epsfbox{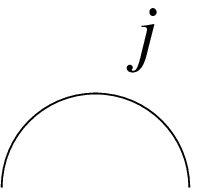} $$
with
$$ \quad\quad\epsfbox{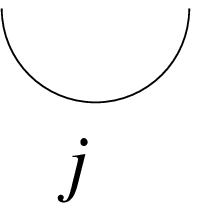} \quad\epsfbox{sn-identityj.eps}  $$
results in
$$\epsfcenter{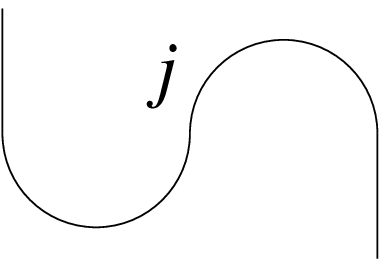}=\epsfcenter{sn-identityj.eps}$$
An example of the tensor product is
$$ \epsfcenter{sn-minj.eps} \otimes \epsfcenter{sn-identityj.eps}\quad = \epsfcenter{sn-minj.eps} \epsfcenter{sn-identityj.eps}$$

The representations and the canonical intertwining operators are built out of the spin $1/2$ representation and a few basic intertwiners, namely the identity, the max, min and crossing diagrams
$$\epsfbox{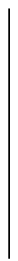}\hskip.5in\epsfbox{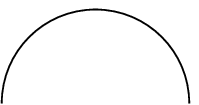}\hskip.5in\epsfbox{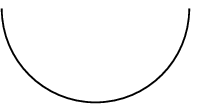}\hskip.5in\epsfbox{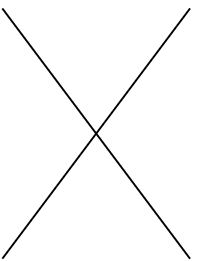}$$
The convention used here is that any line without a label is spin $1/2$.

In fact one does not need to know the tensor representation of the intertwining operators; one can work entirely with these basic diagrams and the relations between them. This point of view is developed systematically in \cite{KL} for the more general $q$-deformed theory, as used in the Turaev-Viro model; the results required here can be obtained by specialising all formulae to $q=1$.  In fact Kauffman introduced a parameter $A$ as the deformation parameter, with $A^2=q$. This means that the formulae in \cite{KL} can be specialised to the Ponzano-Regge case by choosing either $A=1$ or $A=-1$. This choice will be explained below.

The basic relations for spin $1/2$, with $A=\pm1$, are
\begin{equation}\label{loop} \epsfcenter{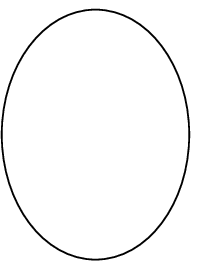} =-2.\end{equation}
and
\begin{equation}\label{skein} \epsfcenter{sn-crossing.eps}=A\;\left(\quad\epsfcenter{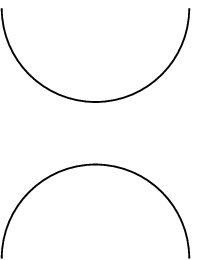}\quad+\quad\epsfcenter{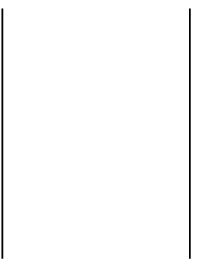}\quad\right)\end{equation}

These relations involve two different signs which need explanation, the minus sign in (\ref{loop}), and the choice of $A=\pm1$ in (\ref{skein}). These are explained in the next two sections.

\subsubsection{The sign in the spherical category}
The minus sign in (\ref{loop}) is the inevitable consequence of the relation
$$\epsfcenter{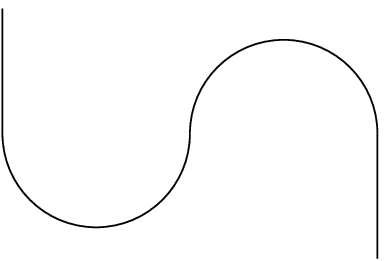}\quad=\quad\epsfcenter{sn-identity.eps}$$
which fixes the tensor representation of the max diagram given the tensor definition of the min.
Invariance under $\SU(2)$ means that the min is an antisymmetric tensor, $\epsilon_{ij}$, on two-dimensional spin-space, but the overall scaling is arbitrary. The loop in (\ref{loop}) has value $\ell=\epsilon_{ij}\epsilon^{ij}$, where $\epsilon_{ij}\epsilon^{jk}=\delta_i^k$. Thus $\ell$ is trace of \emph{minus} the identity, which is $-2$.

This relation generalises to the case of a loop labelled with the spin $j$ irreducible. In this case, the value of the diagram is the quantum dimension.
$$ \epsfcenter{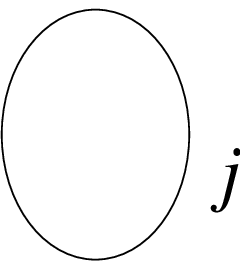}\;=\;(-1)^{2j}(2j+1).$$
This minus sign for the odd spin representations can be understood more abstractly from the point of view of spherical categories. For any group one can construct the category of representations. To define consistently the max and min diagrams with the correct relations, one needs a little extra structure which makes the representation category a \emph{spherical category} \cite{BWS}. For $\SU(2)$ there are exactly two spherical categories, parameterised by the choice of spherical element 
$$w=\pm\begin{pmatrix}1&0\\0&1\end{pmatrix}\in\SU(2).$$
The role of the spherical element is to provide an identification between a representation space $V$ and the dual of its dual, $V^{**}$, the identification being $x\in V$ with $\xi\in V^{**}$, where $\xi(\phi)=\phi(wx)$, for any $\phi\in V^*$.
The value of the loop diagram is then $\tr_j w$, the trace of $w$ in the spin $j$ representation. So the choice of $w=-\begin{pmatrix}1&0\\0&1\end{pmatrix}$ gives the spherical category which is used in the Ponzano-Regge model.

What of the more obvious choice $w=\begin{pmatrix}1&0\\0&1\end{pmatrix}$, leading to the quantum dimension being the obvious classical dimension $2j+1$? There is a state sum model for this category, as there is for any spherical category \cite{BWPL}. However one has to distinguish carefully in the formalism between a representation and its dual. Unlike the Ponzano-Regge case, one cannot \emph{coherently} identify a representation and its dual, in the sense of \cite{BWS}. This means that the spin networks require directed edges to give an unambiguous definition. Moreover, the $6j$-symbols for this category are not symmetrical under the permutation of the vertices of the tetrahedron. One must use the general definition of state sum of a spherical category which requires the vertices in the manifold to be ordered and the manifold oriented. This state sum is manifestly not the Ponzano-Regge model. As noted in \cite{BWPL,MU}, there is also a problem with constructing a vector space of quantum states for a boundary surface in this case.

\subsubsection{The sign in the crossing}

The Ponzano-Regge model is actually independent of the choice of $A=\pm1$ in (\ref{skein}). This is because the spin networks required in the definition of the Ponzano-Regge model are all planar (i.e., without crossings), and also the canonical choice of vertex intertwiners can also be expressed without using crossings. In fact the whole notion of a spherical category is defined only for planar diagrams. 

However, it is useful to use diagrams with crossings as intermediate steps in calculations, and for this the R-matrix provided by (\ref{skein}) is used. Calculating the right-hand side of (\ref{skein}) gives the intertwiner 
$$A\left(\epsilon_{ij}\epsilon^{kl}+\delta^k_i\delta^l_j\right)=A\delta^k_j\delta^l_i
$$
For $A=1$ this is just the flip map $\xi\otimes\eta\mapsto \eta\otimes\xi$ and the spin networks are the spinor calculus. For $A=-1$, the crossing diagram is minus the flip map, which is a fermionic version of the spinor calculus called the binor calculus, first introduced by Penrose \cite{P}. The relation between these two calculi is explained in \cite{BS}. Although either can be used, the binor calculus proves to be more flexible and therefore $A=-1$ is used in the rest of the paper.

It is worth noting that the use of diagrams in spin network calculations is essential. For example, the value of the theta network depends, in general, on the diagram. A calculation shows that
$$ \epsfcenter{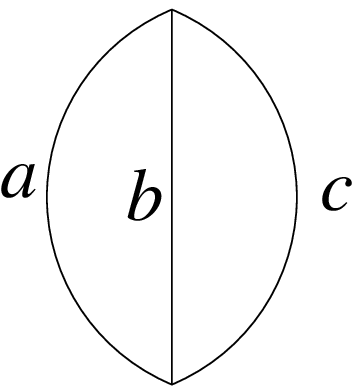}\quad \ne \quad \epsfcenter{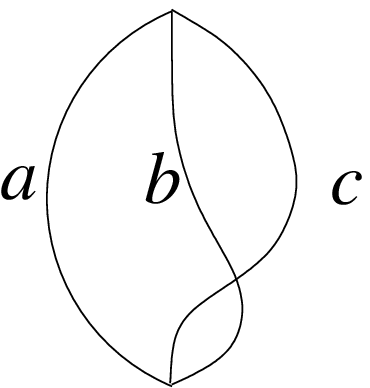}.$$
This means that the spin network value cannot be regarded as a property merely of the one-dimensional spin network graph. 

\subsubsection{The weight}

The Ponzano-Regge weight can be written in a more conceptual way using spin networks. Defining $\left[\begin{matrix}j_1&j_2&j_3\\j_4&j_5&j_6\end{matrix}\right]$ by the tetrahedral spin network
$$\epsfcenter{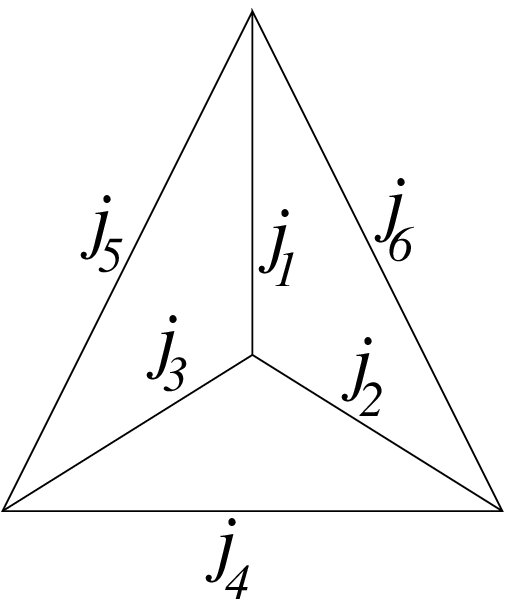}$$
 and $\theta(a,b,c)$ by the theta network
$$ \epsfcenter{thetanetwork.eps}$$
the $6j$-symbol is given by
$$\left\{\begin{matrix}j_1&j_2&j_3\\j_4&j_5&j_6\end{matrix}\right\}=
\frac{\left[\begin{matrix}j_1&j_2&j_3\\j_4&j_5&j_6\end{matrix}\right]}{\sqrt{|\theta(j_1,j_2,j_3) \theta(j_2,j_4,j_6)\theta(j_3,j_4,j_5)\theta(j_1,j_5,j_6)|}},$$
as follows from comparing the formulae in \cite{KL} with the definition of the $6j$-symbol. Since the sign of $\theta(j_1,j_2,j_3)$ is equal to $(-1)^{j_1+j_2+j_3}$, the weight for a closed manifold can be written
\begin{equation}\label{conceptual}
W=\prod_{\text{edges}}(-1)^{2j}(2j+1) \prod_{\text{triangles}}\theta(j_1,j_2,j_3)^{-1}\prod_{\text{tetrahedra}}\left[\begin{matrix}j_1&j_2&j_3\\j_4&j_5&j_6\end{matrix}\right].
\end{equation}

\subsection{Introducing group variables}

In this section it is shown how to rewrite the Ponzano-Regge weight as an integral over group elements on dual edges. The reason for doing this is that it allows the regularisation of the Ponzano-Regge model to be carried out by regularising the integrals. The main idea of the argument is given in \cite{O,BO,KNS}, though the details of the sign factors given here are new. Curiously, it is necessary to restrict to the case where the 3-manifold is orientable. We do not know if an analogous construction works in the non-orientable case. For simplicity of presentation the result is restricted to closed manifolds, though boundaries could be incorporated in a straightforward way.

The group elements are introduced as follows (see also section \ref{group}). There is one variable $g\in\SU(2)$ for each triangle in the manifold, and depends on a choice of orientation (direction) of the edge dual to the triangle in the dual complex (i.e. a choice of normal direction). The element $g^{-1}$ corresponds to the opposite choice of orientation. The product of the $g$ in one complete circuit around the edge is the holonomy of the edge $h$, and is well-defined up to conjugation.

The overall result is as follows. 
\begin{theorem} The weight (\ref{weight}) for an oriented closed manifold can be written
\begin{equation}\label{firstorder}
W=\prod_{\text{triangles}}\int \dd g \prod_{\text{edges}}(2j+1)\Tr_j(h).
\end{equation}
\end{theorem}

Note there is {\em no} factor of $(-1)^{2j}$.

The proof of this result proceeds using the standard identity
\begin{equation}\label{fusing} \epsfcenter{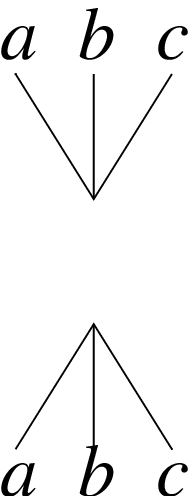} = \theta(a,b,c)\int_{\SU(2)}\;\dd g \quad \epsfcenter{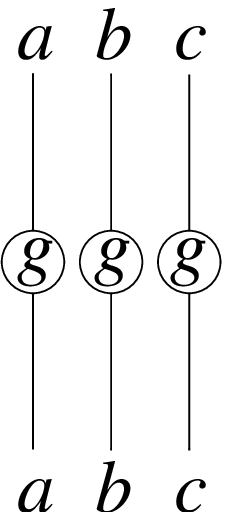} \end{equation}
assuming $\int\dd g=1$. In this identity the strings containing $g$ stand for the matrix representation of $g$ in the representation labelling the string. 

\begin{figure}
$$\epsfcenter{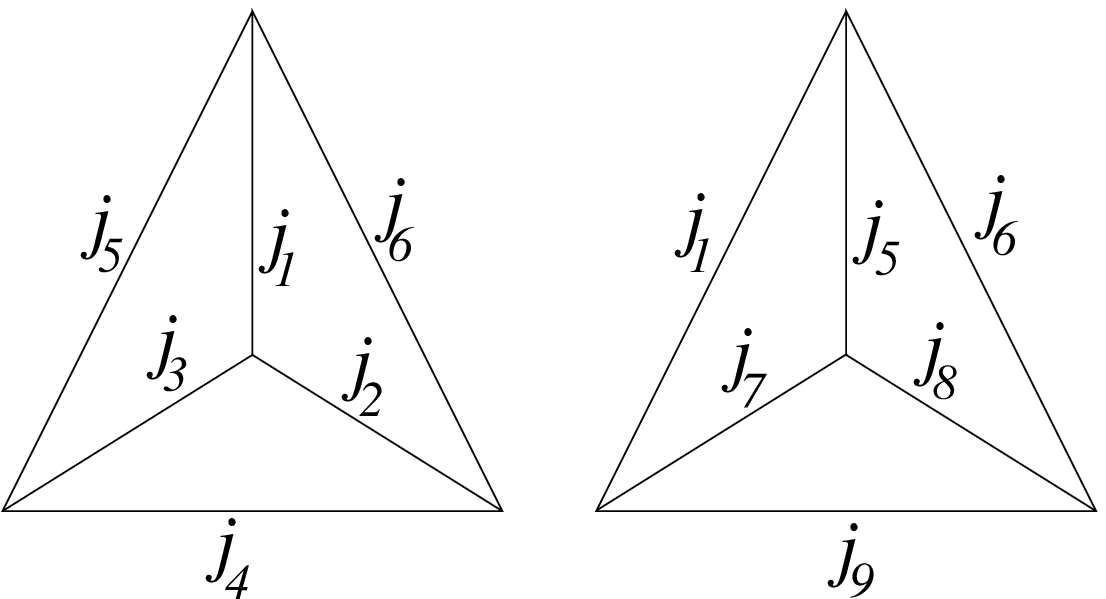}\quad\quad=\quad\theta\int\dd g\quad\epsfcenter{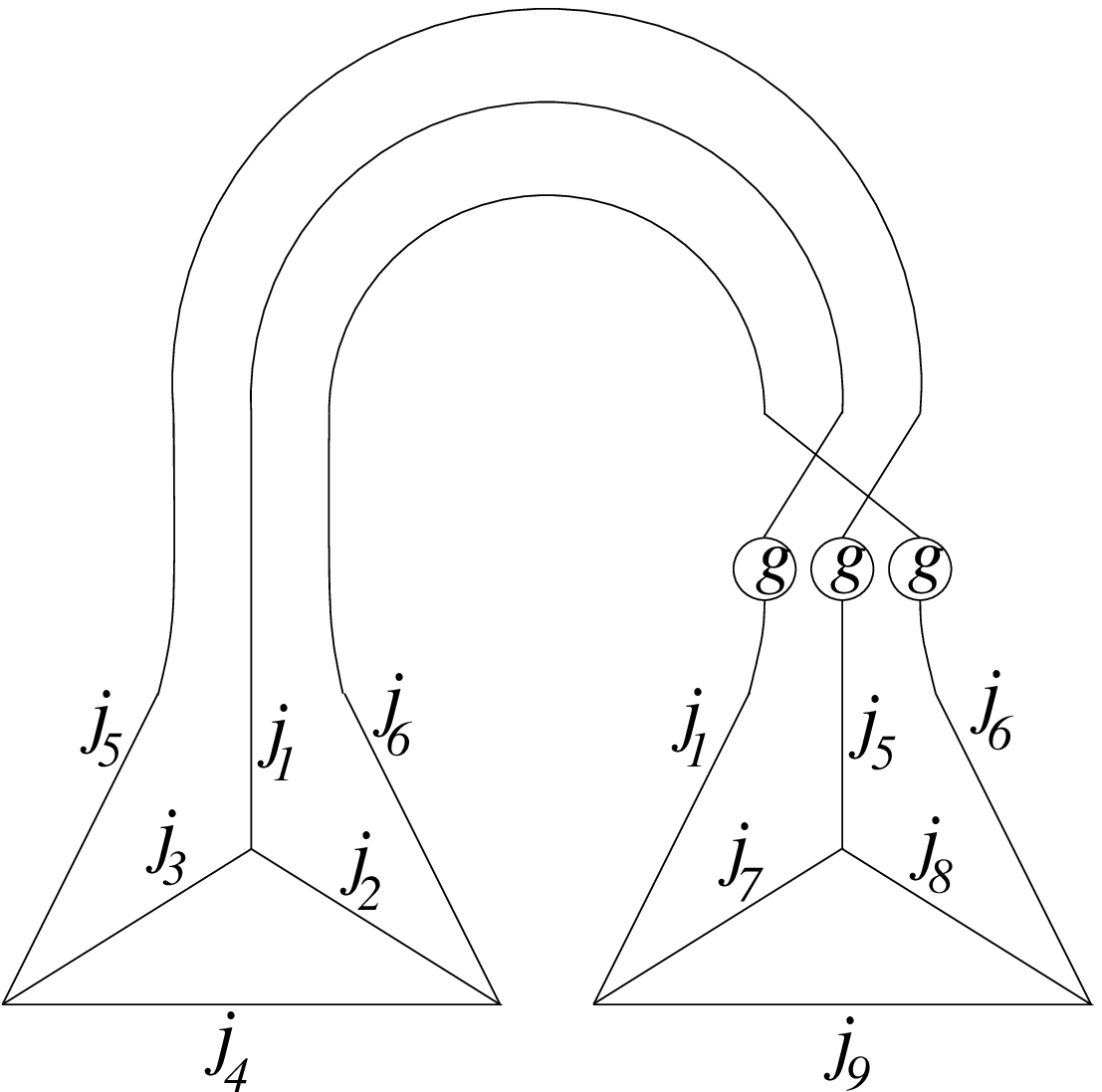}$$
\caption{Fusing two oriented tetrahedra with an even permutation}\label{fusingtets}\end{figure}

\begin{figure}$$\epsfcenter{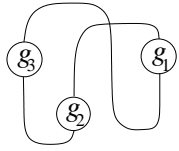}$$\caption{Example of one of the loops after fusion}\label{afterfusion}\end{figure}

The identity is applied to the product of the tetrahedral spin networks in (\ref{conceptual}). 
Each tetrahedron is realised as a spin network in the planar diagram with the orientation of the spin network determined by the orientation of the tetrahedron in the manifold. This means that the cyclic order of the edges meeting at a vertex of the spin network is determined by the orientation of the corresponding triangle in the boundary of the tetrahedron. In an oriented manifold a given triangle appears with one orientation in one tetrahedron and the opposite orientation in the other tetrahedron to which it belongs. This ensures that the permutation of the three edges on the trivalent vertices is an {\em even} permutation, as in figure \ref{fusingtets}.
The relevant modification of (\ref{fusing}) to take into account the permutation of the edges is
$$\theta\int\dd g\quad\epsfcenter{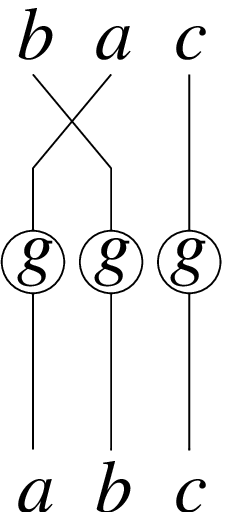}\quad\quad=
\epsfcenter{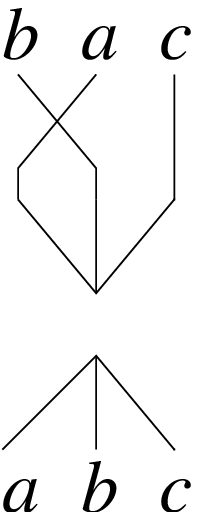}\quad\quad=
\quad(-1)^{a+b-c}A^{4ab}\quad \epsfcenter{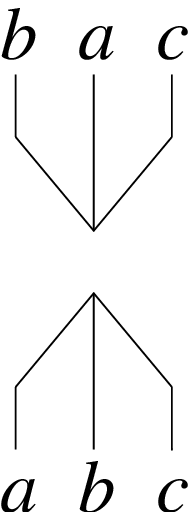}$$
Using this identity twice gives
$$\theta\int\dd g\quad\epsfcenter{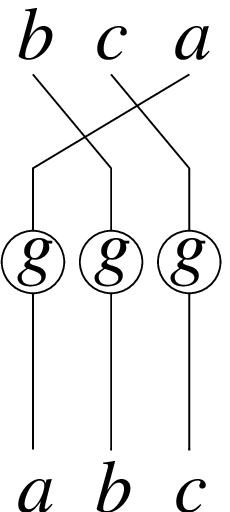}\quad\quad=\quad(-1)^{2a} A^{2a} \quad \epsfcenter{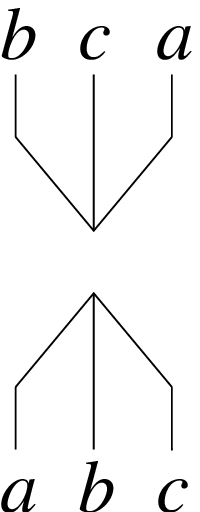}$$
for this even permutation.
Hence taking the binor calculus, $A=-1$, gives no factor of $-1$ on the right hand side. Therefore in the case of an orientable manifold, the tetrahedra can be fused pairwise on all faces to give a spin network diagram which has a number of closed loops, each with a number of insertions of elements $g$ of $\SU(2)$, with no additional factors of $-1$. These loops can be separated to give a product of diagrams such as figure \ref{afterfusion}.
The loop can be simplified in two ways. First, the group elements can be moved together using the identity
$$ \epsfcenter{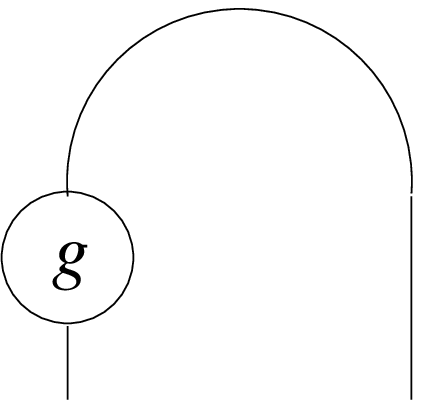} \quad = \quad  \epsfcenter{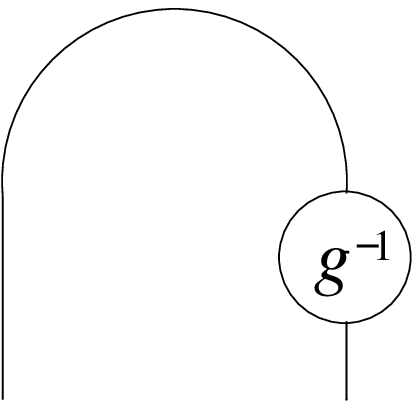} $$
and a similar identity for the min diagram.
Secondly, any self-intersection of the loop can be removed, since the binor calculus satifies the first Reidemeister move.
$$\epsfcenter{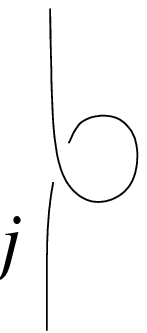}\quad\quad=\quad\quad\epsfcenter{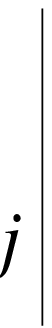}$$
Therefore each loop evaluates to $(-1)^{2j}\Tr_j(h)$ and
$$\prod_{\text{triangles}}\theta(j_1,j_2,j_3)^{-1}\prod_{\text{tetrahedra}}\left[\begin{matrix}j_1&j_2&j_3\\j_4&j_5&j_6\end{matrix}\right]=\prod_{\text{edges}}
(-1)^{2j}\Tr_j(h).$$
The proof of the theorem follows by substitution in (\ref{conceptual}).

As a final remark, one can see why the proof fails for a non-orientable manifold. Any choice of orientation for the tetrahedra leads to at least one triangle on which the tetrahedra are glued with an orientation-preserving map. This means that the corresponding spin networks are fused with an odd permutation. In that case the factor $(-1)^{a+b+c}A^{4ab}$ is not always $1$, and so some extra sign factors spoil the formula (\ref{firstorder}) for the weight.  

\subsubsection{Summing over spins}
Using (\ref{firstorder}) in the formula (\ref{PR}) for the partition function gives an expression which is of course still divergent in some cases. However it is now possible to view the divergence as due to either the sum over spins or the integral over the group variables. In essence the formula is a discrete version of the first-order formalism for quantum gravity.

Summing over the spin labels can be carried out using the identity
\begin{equation}\sum_j(2j+1)\Tr_j(h)=\delta(h)\label{delta}\end{equation}
giving the delta function at the origin of the group (as defined by \eqref{deltagroup}). This gives a divergent formula for the partition function which is written entirely in terms of an integral over variables in $\SU(2)$
\begin{equation} Z=\prod_{\text{triangles}}\int\dd g\prod_{\text{edges}}\delta(h).
\label{divergent}\end{equation}
The regularisation of this formula is the subject of the next section.

Observables in the Ponzano-Regge model can be inserted by using arbitrary functions of the spin variables in place of the factor $2j+1$. It is convenient to use the set of functions given by the character formula \cite{FL3}
$$\frac{\sin\left((2j+1)\frac\theta2\right)}{\sin\frac\theta 2}$$
Inserting such an observable in the sum over spin labels (\ref{delta}) gives
\begin{equation}\label{characterdelta}\sum_j \frac {\sin\left((2j+1)\frac\theta2\right)} {\sin\frac\theta 2}  \Tr_j(h)=\frac{\pi\delta(\theta-c(h))}{\sin^2(\theta/2)}\end{equation}
with $c(h)$ denoting the conjugacy class of $h$ parameterised by the angle of the corresponding rotation, i.e., $c(h)\in[0,2\pi]$.

Putting $\theta=0$ in the left-hand side of (\ref{characterdelta}) shows that the formula reduces to (\ref{delta}). However, the right-hand side of (\ref{characterdelta}) looks singular in this limit. This is because $\theta$ is not a good coordinate on $\SU(2)$ at the identity. To avoid any potential difficulty arising from this, the observable used in the following is just $\delta(\theta-c(h))$ instead of the right-hand side of (\ref{characterdelta}). This amounts to a scaling by a factor of 
$$\frac{\sin^2(\theta/2)}\pi $$
for each observable. This factor is the volume of the conjugacy class of $\theta$.

\section{Regularisation using integrals}\label{group}
This section develops a definition of the Ponzano-Regge partition function as an integral over variables in $\SU(2)$ by regularising the formula developed in the last section. Observables specified by conjugacy classes in $\SU(2)$ are included.

Let $M$ be a closed 3-manifold with a specified triangulation.  As $M$ is compact, the triangulation will have a finite number of simplexes. To specify an observable, we need a graph embedded in $M$, and some data on each edge of the graph. More precisely, let $\Gamma$ be a connected subcomplex consisting of edges and vertices of $M$. The graph $\Gamma$ is assumed to be non-empty; the smallest possible $\Gamma$ is a single vertex and no edges. The observable is specified by $\Gamma$ together with an angle $\theta_e\in [0,2\pi]$ for each edge $e$ of $\Gamma$.
In the following it is necessary to use a more explicit notation for the dependence of variables on the simplexes of the triangulation than that used in the previous sections.

Let $\Delta_n$ be the set of $n$-simplexes of the triangulation. To define the regularisation of the partition function, it is necessary to pick a subset of edges $T\subset\Delta_1$. This should be a set of edges which are not contained in $\Gamma$. This should satisfy the following conditions:
\begin{itemize}
\item
Each connected component of the graph formed by $T$ is a tree (i.e. contains no loops) and is attached to $\Gamma$ at exactly one vertex
\item $T$ is maximal, i.e. visits each vertex of $M$ not contained in $\Gamma$.
\end{itemize}
 
The definition of the partition function is as follows. We use the dual cell decomposition of $M$ in which there is one dual $k$-cell for each $3-k$--cell of $M$. An oriented dual edge is determined by an ordered pair of neighbouring dual vertices, $f=(v_0,v_1)$, and is regarded as having a direction or arrow from $v_0$ to $v_1$.

On each oriented dual edge $f$ of $M$ there is a variable $g_f\in\SU(2)$. The inverse of $g^{-1}_f$ is assigned to the opposite orientation of $f$. This set of variables is called a connection, and given an oriented path consisting of a sequence of oriented dual edges $\gamma=(f_1, f_2, \ldots, f_N)$, with the orientations of $f_i$ agreeing with the orientation of the path, there is a holonomy element
\begin{equation*}
H(\gamma) \; = \; g_{f_1} \, g_{f_2}\, \ldots \, g_{f_N}.
\end{equation*}
 
On each oriented dual face $e$, there is then the  holonomy $h_e=H(\gamma)$ given by the sequence $\gamma$ of dual edges around its boundary. This is well-defined up to conjugation. Finally, the definition uses some delta-functions on $\SU(2)$. The first of these is the three-dimensional delta-function at the identity element $\identity$, defined by
\begin{equation}\label{deltagroup}\int_{\SU(2)}\delta(g)F(g)\dd g=F(\identity),
\end{equation}
for any function $F$, where $\smallint \dd g = 1$. Let $c(g)$ denote the angle of rotation in Euclidean space determined by $g\in\SU(2)$. 
The second is the delta-function at a conjugacy class $\phi$, given by an ordinary one-dimensional delta-function $\delta(\phi-c(g))$.  
 
The set of all variables is indexed by the set $\Delta_2$ of \emph{unoriented} dual edges; picking an arbitrary orientation for each $f\in\Delta_2$ gives a variable $g_f$, and these are all independent.
The partition function is obtained by integrating over these variables. 
\begin{equation}\label{partition}
Z= 
\int\prod_{f\in\Delta_2} \dd g_{f}\prod_{e\in\Gamma}\delta(c(h_e)-\theta_e)\prod_{e'\in\Delta_1\setminus(\Gamma\cup T)}\delta(h_{e'})\end{equation}

A similar definition appears in \cite{FL1,FL2,FL3}, but with a somewhat different definition of $T$.\footnote{For example, if $\Gamma$ includes every vertex of the triangulation but not every edge, then $T$ defined here is empty, but the analogous set defined in \cite{FL1} is not.}

It is clear that the delta-functions for the edges on $\Gamma$ force the holonomy of the connection around that edge of the graph to lie in the conjugacy class $\theta_e$.
The remainder of this section shows that the effect of the delta-functions at the identity is to force the $g$ variables to give a flat $\SU(2)$ connection on the complement of $\Gamma$.  The role of the set of edges $T$ is to eliminate excess delta-functions, which would otherwise reduce to integrating the square of a delta-function in one of the variables. 

That $T$ should be maximal is clear, for suppose $T$ is not maximal and consider what happens around a vertex not in either $\Gamma$ or $T$. We would write down a delta-function $\delta\left(h_e\right)$ for every edge $e$ incident on that vertex. But the holonomy around any one of those edges may be expressed as the product of the holonomies around all the remaining edges, giving us one too many delta-functions. 

It is less obvious that, with the given definition of $T$, there are sufficiently many delta-functions to force a flat connection on the complement of $\Gamma$. That is, it remains to prove the following

\begin{lemma} If $h_e = \identity \; \forall \; e \in \Delta_1\setminus(\Gamma\cup T)$ then $h_e = \identity \; \forall  \; e \in \Delta_1\setminus\Gamma$.
\end{lemma}

\begin{figure}[htb]
$$\epsfbox{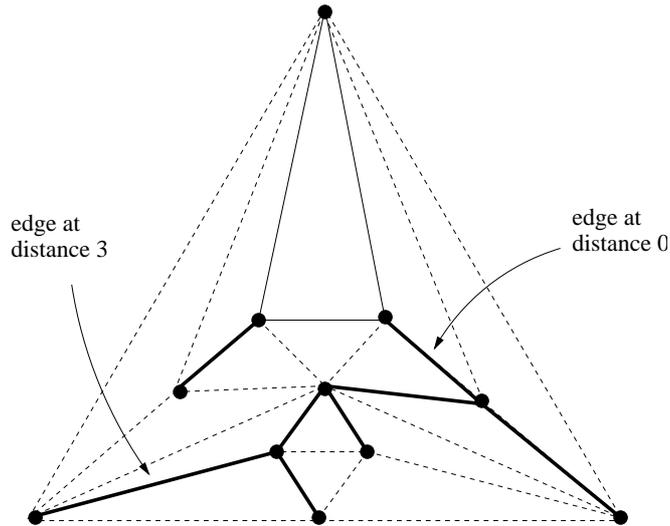}$$
 \caption{A triangulation (dashed lines) of ${S}^2$ containing the unknot (continuous lines). The regularising edges (bold lines) have $N = 4$.}
 \label{distance}
\end{figure}

\noindent\emph{Proof.}
By induction on the edges of $T$. First, orient each edge of $T$ to point away from $\Gamma$ (so that we may refer to final and initial vertices). Define the \emph{distance} of an edge $e \in T$ from $\Gamma$ to be the number of edges traversed in travelling from the base of the tree containing $e$ to the initial vertex of $e$ (along edges of $T$) (see figure \ref{distance}). Let $N-1$ be the maximum distance of an edge of $T$ from $\Gamma$ (so for example in figure \ref{distance}, $N=4$). Define the statement 
\begin{equation*}
P_i: \; h_e = \identity \; \forall \; e \in T \text{ at a distance } N-i \text{ from } \Gamma. 
\end{equation*}
If we can prove that for $i \in \lbrace1,\ldots, N\rbrace, \: P_{i-1} \Rightarrow P_i$, we will be done, since $P_0$ is vacuous. So, let $i \in \lbrace1,\ldots, N\rbrace$ and assume $P_{i-1}$. Consider an edge $e \in T$ at a distance $N-i$ with final vertex $v$. For each edge $e' \neq e$ incident on $v$, we have $h_{e'} = \identity$. But, as we have argued previously, this situation implies $h_e = \identity$. Finally, since $e$ was arbitrary, $P_i$ is true.
 
\section{The existence of the partition function}\label{EPF}
Even with the regularisation, the partition function is not always well-defined.
A criterion which distinguishes the manifolds and observables for which the partition function is well-defined is presented in this section.

{\def\epsfsize#1#2{0.25#1}
\begin{figure}[htb]
$$\epsfbox{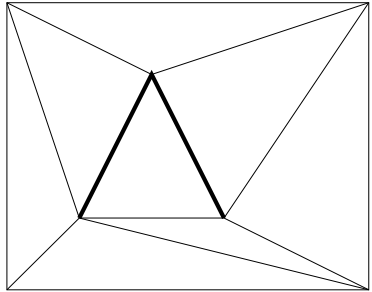}$$
 \caption{An example of $\Gamma$}
 \label{rn-triangulation}
\end{figure}
\begin{figure}[htb]
$$\epsfbox{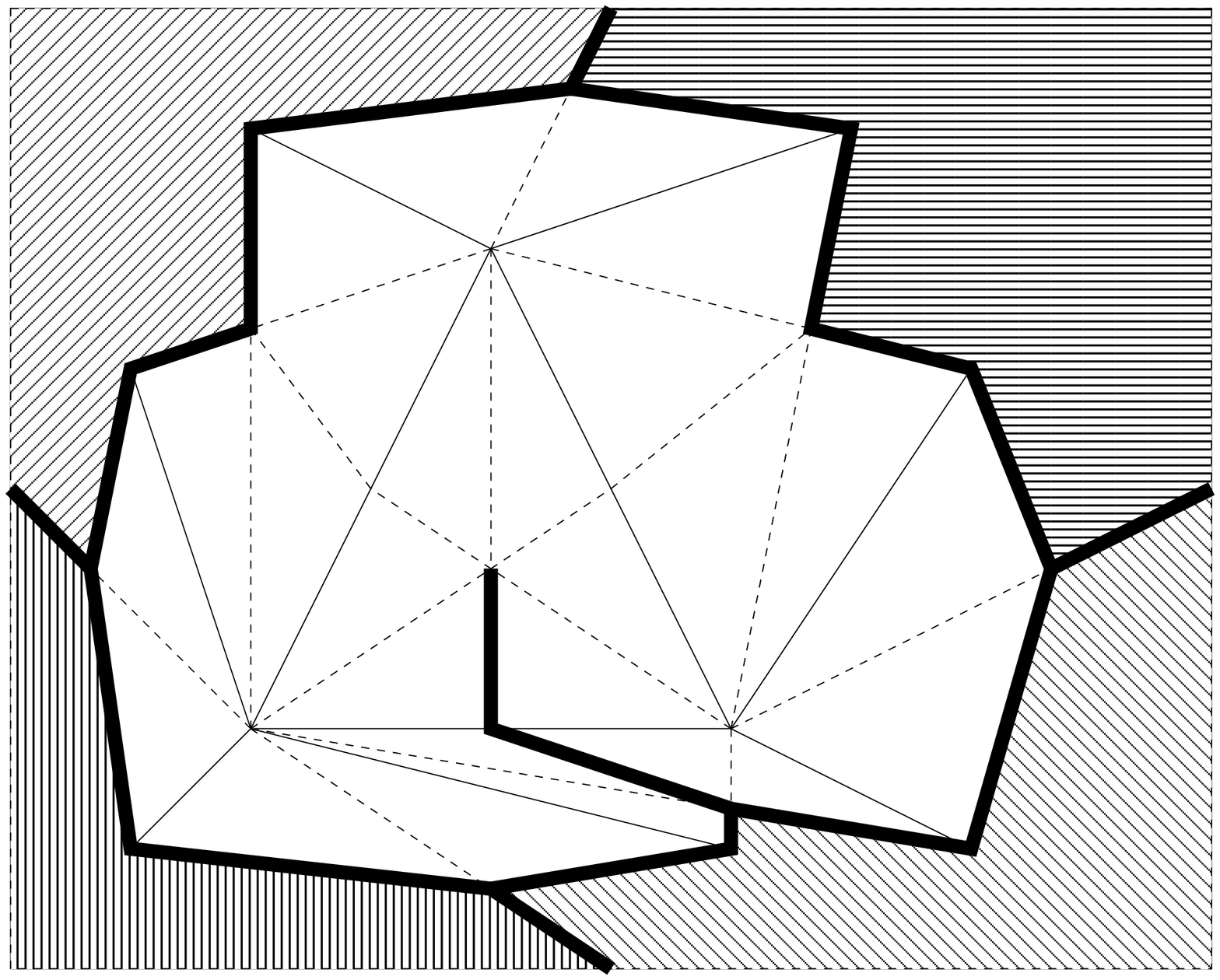}$$
 \caption{The cell complex $L$ for the example of figure \ref{rn-triangulation}. The 2-cells are shaded and the 1-cells are in bold.}
 \label{rn-dual}
\end{figure}
}

\subsection{Exterior complexes}\label{collapsing}
Two subcomplexes of the dual complex of $M$ play a key role in the following.
The cell complex $L$ is obtained from the dual cell complex of $M$ by removing the dual 3-cell to each vertex of the graph $\Gamma$ and the 2-cell dual to each edge of $\Gamma$. In general $L$ is a 3-complex (all cells have dimension $\le3$). In many cases $L$ is a manifold with boundary, but it can happen that $L$ is not a manifold. See for example figure \ref{rn-triangulation} and figure \ref{rn-dual} with $M$ a 2-manifold and $L$ a 2-complex. (The dimension has been reduced to make it possible to draw the figures.) The dual cells are constructed as a union of simplexes of the barycentric subdivision of $M$. The barycentric subdivision is indicated with dotted lines in figure \ref{rn-dual}, and the 2-cells that have been removed to form $L$ each consist of the simplexes of the barycentric subdivision which contain the corresponding vertex of $\Gamma$.

The other cell complex needed in the following is the subcomplex of $L$ obtained by removing all 3-cells and all 2-cells dual to the edges of $T$. This is denoted $K$ and is a 2-complex.
The complex $K$ is obtained from $L$ by the process of collapsing, which is now explained.
\begin{figure}[htb]
$$\epsfbox{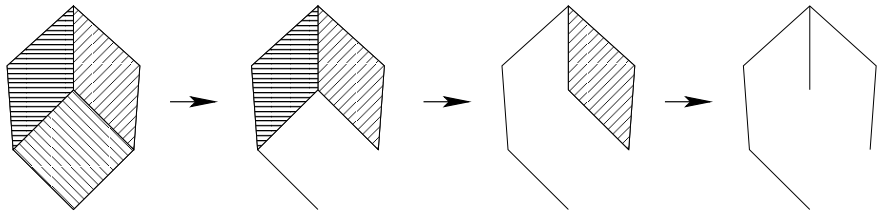}$$
 \caption{A collapsing in two dimensions}
 \label{collapse}
\end{figure}

Collapsing is an operation on a cellular complex $C$ in which there is a $k$-cell $\tau$ which appears in the boundary of only one $k+1$-cell $\sigma$. Then the operation is to remove both $\sigma$ and $\tau$ from the cell complex. For example, if $C$ is a connected 3-manifold with boundary, then any of the three-cells which meet the boundary of $C$ in a disk are suitable for collapsing. The collapsing can be repeated until there are no 3-cells left. This is called collapsing $C$ to a 2-skeleton.

In our case, the 3-cells are the duals of the vertices in the triangulation. A sequence of collapsings $L\to K$ of $L$ to a 2-skeleton $K$ is a sequence of edges $T$ of the triangulation (each dual to the 2-cell $\tau$ across which the collapsing takes place). The fact that collapsing occurs across a 2-cellular face which is either on the original boundary or on a face exposed by a previous collapse leads, in the dual picture, to exactly the conditions of a regularising subset of edges of section \ref{group}.   This equivalence of a collapsing to a 2-skeleton with a regularising subset of edges explains why this is a good definition of $T$ (see also section \ref{invariance}).
\begin{figure}[htb]
$$\epsfbox{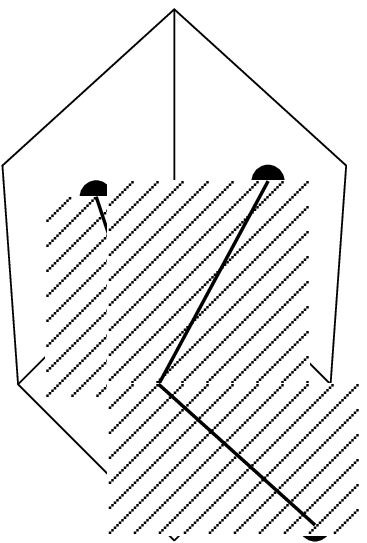}$$
 \caption{The tree $T$ for the collapsing of Figure \ref{collapse}.}
\end{figure}

\subsection{The exterior manifold}
Examples show that the topology of $L$ depends on the triangulation of $M$. This section discusses how $L$ can be replaced by a canonically defined manifold $T(L)$ called the exterior manifold of the graph $\Gamma$ in $M$.

For example, in figure \ref{rn-dual}, $L$ is not a manifold. However, if the central triangle together with the triangle below it in figure \ref{rn-triangulation} were subdivided suitably, then it would be. (This example is two-dimensional but the principle is the same.)
\begin{figure}[htb]
$$\epsfbox{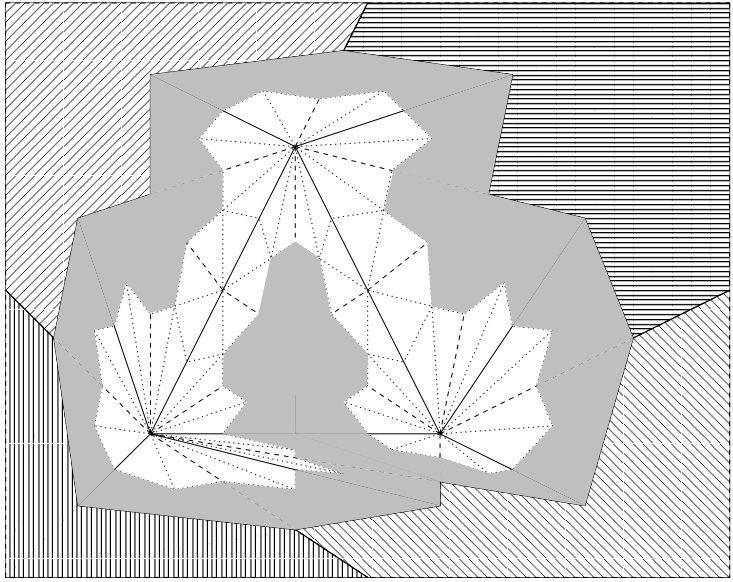}$$
 \caption{The thickening of $L$ to the manifold $T(L)$ by a regular neighbourhood. The additional cells are solid grey. The remaining white area is the regular neighbourhood of $\Gamma$}\label{rn-detail}
\end{figure}
This dependence on the triangulation can be avoided by the construction of a regular neighbourhood of $L$ in $M$ \cite{RS}. The regular neighbourhood of $L$, denoted $T(L)$, can be thought of as a thickening of $L$. This is a canonical construction determined by subdividing the triangulation around $L$ a second time (see figure \ref{rn-detail}, which shows the construction for the complex in figure \ref{rn-dual}). According to the regular neighbourhood theorem, a regular neighbourhood of $L\subset M$ is unique up to isomorphism of $M$ that fixes $L$, i.e. it is independent of the triangulation. In fact the complement of $T(L)$, i.e. $M\setminus \mathrm{interior}(T(L))$, is a regular neighbourhood of the original graph $\Gamma$. This shows that $T(L)$ is uniquely determined by $\Gamma\subset M$ and is independent of the triangulation of $M$. Both $T(L)$ and $T(\Gamma)=M\setminus \mathrm{interior}(T(L))$ are manifolds, and they meet on their boundary. Moreover $T(\Gamma)$ collapses to $\Gamma$ and $T(L)$ collapses to a triangulation of $L$. This means that any topological property of $L$ which is invariant under collapsing is a property of $T(L)$ and hence canonical. 

\subsection{Twisted cohomology}\label{twistedcohomology}
The set of variables $g_f$, subject to the relations $h_e=\identity$ for each dual face $e$, is called a flat connection, and will be denoted by $\rho$. A flat connection
determines a twisted cohomology for the graph exterior. This cohomology is important for understanding the partition function. For this purpose, it is easily computed using the dual cell decomposition, though one obtains the same cohomology with any cell decomposition of $L$. 

The twisted cohomology for the cell complex $L$ can be described as follows.
Given two vertices $v_0$ and $v_1$ of a cell $\sigma$ of $L$, there is a holonomy element $h_{v_0,v_1}=H(\gamma_{v_0,v_1})$, where $\gamma_{v_0,v_1}$ is any path of edges of $\sigma$ from $v_0$ to $v_1$. Since the connection is flat, then it does not matter which sequence of edges within $\sigma$ is used.

In twisted cohomology it is necessary for every oriented cell $\sigma$ to have an (arbitrarily chosen) basepoint $v(\sigma)$, one of its vertices. A $k$-cochain $\Phi\in C^k(L)$ for the twisted cohomology assigns an element of the Lie algebra $\su(2)$ to each oriented $k$-cell $\sigma$. If $-\sigma$ is the same $k$-cell but with the opposite orientation, then $\Phi(-\sigma)=-\Phi(\sigma)$.

The boundary map is twisted by the flat connection. This twisting uses the adjoint action of the group $\SU(2)$ on the Lie algebra $\su(2)$. This action is written $g\action l$, and in terms of matrix multiplication is defined by
$$g\action l=glg^{-1}.$$

If the boundary of $\sigma$ is written as a sum of oriented $k$-cells
$$\partial\sigma=\sum\tau,$$
then the boundary of a cochain is the $k+1$-cochain
\begin{equation}\label{coboundary}
\dd \Phi(\sigma)= \sum h_{v(\sigma),v(\tau)}\action \Phi(\tau).
\end{equation}

The sequence of chain groups is thus
\begin{equation} 0\to C^0(L) \xrightarrow{d_0} C^1(L) \xrightarrow{d_1} C^2(L)\xrightarrow{d_2} C^3(L)\to 0.\label{sequence}\end{equation}  
and the twisted cohomology $H^*(L,\rho)$ is the homology of this sequence. Appendix 1 shows that this definition is equivalent to the one which is usually given in terms of covering spaces.

\subsection{Existence criterion}\label{existence}

Now it is possible to state the main existence result. This says, roughly, that the partition function exists whenever $H^2(L,\rho)=0$. There are two factors which make this complicated. The first is that, for a given graph, the vanishing of this cohomology group may depend on the flat connection $\rho$ and therefore on the values of the parameters $\theta_e$. Secondly, examples (see below) typically give partition functions which have delta-functions in these parameters, so it often does not make sense to speak of the partition function for specific values of the parameters. Hence the main result is phrased in terms of a region of parameters.

\begin{theorem} The partition function (\ref{partition}) exists for a region $\cal R$ of the space of parameters $\{(\theta_1,\theta_2,\ldots)\}$ as a distribution if and only if the second twisted cohomology group $H^2(L,\rho)$ is trivial for each flat connection $\rho$ whose conjugacy classes  for holonomy around the edges of $\Gamma$, $(c(h_1),c(h_2),\ldots)$, lie in $\cal R$. The same result holds if $L$ is replaced by the graph exterior $T(L)$.
\end{theorem}

\noindent\emph{Proof.}
Let $\psi(\theta_1,\theta_2,\ldots)$ be a continuous test function. Then integrating it with $Z$ defined in (\ref{partition}) gives
\begin{equation*}
\int Z\psi\; \prod\dd\theta\;= 
\int\prod_{f\in\Delta_2} \dd g_{f}\;\psi(c(h_1),c(h_2),\ldots)\prod_{e'\in\Delta_1\setminus(\Gamma\cup T)}\delta(h_{e'})\end{equation*}
This formula involves the integration of the delta function $\delta(F)$, where $F$ is the map
 $$F(g_1,g_2\ldots,g_m)=(h_1,h_2,\ldots,h_n).$$
On the right-hand side, there is one holonomy element for each 2-cell in $K$, the 2-skeleton of $L$ obtained by collapsing.
The delta-function is well-defined if the differential $F_*$ of the map is a surjection. In this case the integral exists. Hence to prove the theorem, it remains to show that $F_*$ is a surjection.  

This differential can be expressed in terms of the twisted cohomology. This is done by establishing isomorphisms between the tangent spaces to $\curly M=\SU(2)^m$ and $\curly N=\SU(2)^n$ and the vector spaces of twisted cochains. Then it will be shown that $F_*$ becomes equal to the twisted boundary map $\dd_1$.

It is convenient to use a derivative notation for tangent vectors, representing the tangent vector as a derivative along a path. The isomorphism $\Xi_1 \colon C^1(K)\to{T\curly M}_\rho$, with $\rho=(g_1,g_2\ldots,g_m)$, maps a cochain $\Phi$ to the tangent vector $w=\bigl(\frac{\dd g_1}{\dd t},\frac{\dd g_2}{\dd t},\ldots,\frac{\dd g_m}{\dd t}\bigr)$ with components defined by
$$\frac{\dd g}{\dd t}g^{-1}=\Phi(\sigma)$$
for each oriented 1-cell $\sigma$ with holonomy $g$ and distinguished vertex $v(\sigma)$ at the `from' end.

If, alternatively, a 1-cell $\sigma$ has its distinguished vertex $v(\sigma)$ at the `to' end,
then the formula is instead
$$g^{-1}\frac{\dd g}{\dd t}=\Phi(\sigma).$$

The corresponding vector in $\curly N$, $F_*(w)\in {T\curly N}_{F(\rho)}$, has components given by the following calculation. Let $\sigma'$ be an oriented 2-cell with holonomy $h=g_1g_2\ldots g_k$ around its boundary (in the direction determined by the orientation). Then its component of $F_*(w)$ is given by
$$\frac{\dd h}{\dd t}=
\frac{\dd g_1}{\dd t}g_2\ldots g_k+ g_1\frac{\dd g_2}{\dd t}\ldots g_k+\ldots + g_1g_2\ldots\frac{\dd g_k}{\dd t}.$$
Hence
\begin{equation}\label{Df}
\frac{\dd h}{\dd t}h^{-1}=\frac{\dd g_1}{\dd t}g_1^{-1}
+ g_1\action\left(\frac{\dd g_2}{\dd t}g_2^{-1}\right)
+\ldots 
+ g_1g_2\ldots g_{k-1}\action\left(\frac{\dd g_k}{\dd t}g_k^{-1}\right).\end{equation}
Therefore if the isomorphism $\Xi_2 \colon C^2(K)\to T\curly N_{F(\rho)}$ is given by
mapping the cochain $\Phi'$ to the tangent vector $\bigl(\frac{\dd h_1}{\dd t},\frac{\dd h_2}{\dd t},\ldots,\frac{\dd h_m}{\dd t}\bigr)$ given by
$$\Phi'(\sigma')=\frac{\dd h}{\dd t}h^{-1},$$
then, using these isomorphisms, and comparing (\ref{Df}) with (\ref{coboundary}), it follows that
$$ F_*\,\Xi_1=\Xi_2\,\dd_1.$$
This coboundary operator is a surjection only when $H^2(K,\rho)=0$. However cohomology is unchanged under collapsing, so $H^2(T(L),\rho)=H^2(L,\rho)=H^2(K,\rho)=0$.

\section{The invariance of the partition function}\label{invariance}

Our strategy for proving the invariance of the partition function under change of triangulation (or more generally, choice of cell complex) of the manifold, and the independence of the choice of the regularising set $T$ lies in showing that the partition function is given by an integral over the space of flat connections with measure given by the Reidemeister torsion. The Reidemeister torsion \cite{DUB} is known to be an invariant of homeomorphisms, which shows that it is independent of the triangulation. It is also invariant under collapsing, which shows that the choice of $T$ is unimportant.

\subsection{Reidemeister torsion}\label{Reidemeistertorsion}
The Reidemeister torsion is a topological invariant which is constructed using the sequence of  chain groups (\ref{sequence}). More recent work on Reidemeister torsion has emphasised the correct calculation of the sign; however since $Z\ge0$, this is not required here and the modulus of the torsion suffices to express $Z$ invariantly.

Let $V$ be a vector space of dimension $d$ and let $\Omega^p(V)$ be the space of $p$-forms on $V$. If $\mathbf{a}=\{a_1,a_2,\ldots,a_d\}$ is a basis of $V$ and $\omega\in\Omega^d(V)$, then define $<\omega,\mathbf{a}>\in\R$, the evaluation of $\omega$ on $\mathbf{a}$, by the following. Express $\omega$ as a product of $1$-forms,
$$\omega=\theta_1\wedge\theta_2\wedge\ldots\wedge\theta_d.$$
Then the definition is
$$<\omega,\mathbf{a}>=\left|\sum_{\text{permutations}\ \sigma}(-1)^{|\sigma|}\theta_1(a_{\sigma(1)})\theta_2(a_{\sigma(2)})\ldots\theta_d(a_{\sigma(d)})\right|.$$
If $\mathbf{b}$ is another basis of $V$, then define
$$[\mathbf{a}/\mathbf{b}]=\frac{<\omega,\mathbf{a}>}{<\omega,\mathbf{b}>}$$
for any non-zero $d$-form $\omega$. It is straightforward to see that if $a_i=\sum_j M_{ij} b_j$ for some matrix $M$, then $[\mathbf{a}/\mathbf{b}]=|\det M|>0$. (Take $\theta$ to be the basis dual to $\mathbf{b}$.)

To define the Reidemeister torsion, pick a basis $\mathbf{l}$ of the Lie algebra $\su(2)$. A basis $\mathbf{c}^k$ of the chain group $C^k(L)$ is determined by taking the functions which map one of the $k$-simplexes to an element of the basis $\mathbf{l}$ and the others to $0$. Next, pick a basis $\mathbf{h}^k$ for each vector space $H^k(L)$, and choose an arbitrary lifting $\mathbf{{\widetilde h}}^k$ of these vectors to a linearly independent set of vectors in $C^k$. Let $\mathbf{b}^k$ be a linearly independent set of vectors in $C^k$ such that $d_k(\mathbf{b}^k)$ is a basis for the image of $d_k$. Then $d_{k-1}(\mathbf{b}^{k-1})\mathbf{{\widetilde h}}^k\mathbf{b}^k$ forms a second basis for $C^k(L)$. Define
\begin{equation}\label{tauk}
\tau_k=[d_{k-1}(\mathbf{b}^{k-1})\mathbf{{\widetilde h}}^k\mathbf{b}^k/\mathbf{c}^k].
\end{equation}
Then the Reidemeister torsion of the 3-complex $L$ is defined to be
$$\tor(L)=\tau_0^{-1}\tau_1\tau_2^{-1}\tau_3.$$
The definition is constructed so that the dependence on each $\mathbf{b}^k$ cancels. The dependence on the triangulation or cell structure of $L$ is subtle, giving the topological invariance mentioned above. It does depend on the explicit choice of $\mathbf{h}^k$ and $\mathbf{l}$, but is independent of the liftings $\mathbf{{\widetilde h}}^k$. The torsion also depends on the flat connection $\rho$ which is used in the definition of the coboundary operators $d_k$.

\subsection{The partition function in terms of Reidemeister torsion}

The delta functions appearing in the formula for the partition function (\ref{partition}) can be understood in the following general way. If $\omega$ is a volume form on a manifold, then $|\omega|$ denotes the positive integration measure determined by $\omega$, for example,
$$|\dd x\wedge\dd y|=\dd x\,\dd y.$$
If $\curly M$ and $\curly N$ are manifolds, $\mu$ is a $\mathrm{dim}(\curly M)$-form on $\curly M$ and $\nu$ is a $\mathrm{dim}(\curly N)$-form on $\curly N$,
$p \in \curly N$ and $F: \curly M \rightarrow \curly N$ is a function whose differential is surjective at $p$, then $\delta_p(F)$ is defined by
\begin{equation}\label{deltafunction}
\int_\curly M \delta_p(F) |\mu| = \int_{\curly L} |\lambda| \, ,
\end{equation}
where $\curly L = F^{-1}(p)$ and $\lambda$ is the form on the manifold $\curly L$ defined by
\begin{equation}\label{lambda}
\mu = \widetilde\lambda \wedge F^*(\nu) \, .
\end{equation}
In this formula, $\widetilde\lambda$ is a section of the bundle of differential forms on $\curly M$ restricted to $\curly L$, and $\lambda$ is the pull-back of $\widetilde\lambda$ to a differential form on $\curly L$.

The first step in understanding the partition function (\ref{partition}) is to write it in the form of (\ref{deltafunction}) and  use (\ref{lambda}) to calculate $\lambda$. In this case we have $\curly M = \SU(2)^m$ and $\curly N = \SU(2)^n$ where $m$ is the number of 1-cells (dual edges) in $K$ and $n$ is the number of 2-cells in $K$,
$F(g_1,\ldots, g_m) = (h_1,\ldots, h_n)$, $p$ is the identity element in $\curly N$, $\nu = \dd h_1 \wedge \ldots \wedge \dd h_n$, the wedge product of $n$ copies of the Haar measure on $\SU(2)$ and 
\begin{equation}\label{mu1}
\mu = \prod_{e^* \in \Delta_2} \dd g_{e^*} \prod_{e \in \Gamma} \delta\left(\theta_e - c(h_e)\right).
\end{equation}
In the formula for $\mu$, the delta functions do not cause any problems, as they contain external parameters. They could be smoothed by a test function, as in section \ref{existence}. It is assumed that the parameters of the delta functions are chosen such that the twisted $H^2$ is equal to zero; thus $\curly L$ is restricted to these values.

From (\ref{mu1}),
\begin{equation}\label{mu2}
<{\Xi_1}^*\mu,\mathbf{c}^1>\; =\; <\dd g,\mathbf{l}>^m \prod_{e \in \Gamma} \delta\left(\theta_e - c(h_e)\right)
\end{equation}
and from (\ref{lambda})
\begin{align*}
<{\Xi_1}^*\mu,d_0(\mathbf{b}^0) \, \mathbf{\tilde{h}}^1 \, \mathbf{b}^1>\; 
&=\; <{\Xi_1}^*\lambda,d_0(\mathbf{b}^0) \, \mathbf{\tilde{h}}^1> \, 
<{\Xi_1}^*F^*(\nu),\mathbf{b}^1>\\ 
&=\; <{\Xi_1}^*\lambda,d_0(\mathbf{b}^0) \, \mathbf{\tilde{h}}^1> \, 
<{\Xi_2}^*\nu,d_1(\mathbf{b}^1)>
\end{align*}
since the derivative of $F$ is the twisted coboundary map $d_1$. This implies
\begin{align*}
<{\Xi_1}^*\mu,\mathbf{c}^1>\tau_1\; &=\; <{\Xi_1}^*\lambda,d_0(\mathbf{b}^0) \, \mathbf{\tilde{h}}^1> \, <{\Xi_2}^*\nu,\mathbf{c}^2> \, \tau_2\; \\
&=\; <{\Xi_1}^*\lambda, d_0(\mathbf{b}^0) \, \mathbf{\tilde{h}}^1 > \, <\dd h,\mathbf{l}>^n \, \tau_2,
\end{align*}
where we picked up factors of $\tau_1$ and $\tau_2$ by changing bases for $C^1$ and $C^2$. Finally, from (\ref{mu2}) we have
\begin{equation}\label{lambda2}
<{\Xi_1}^*\lambda,d_0(\mathbf{b}^0) \, \mathbf{\tilde{h}}^1 >\; =\; \tau_1 \, \tau_2^{-1} \, <\dd g,\mathbf{l}>^{m-n} \prod_{e \in \Gamma} \delta\left(\theta_e - c(h_e)\right).
\end{equation}

The second step is to extract a further factor of $\tau_0^{\scriptscriptstyle{-1}}$, which together with the first two factors on the RHS of (\ref{lambda2}) will make up the torsion
$$\tau_0^{-1}\,\tau_1 \, \tau_2^{-1}=\tor(K)=\tor(L)=\tor(T(L)).$$
 This is done as follows.
Let $l$ be the number of $0$-cells in $K$. The group $\curly G=\SU(2)^l$ acts on $\rho\in\curly L$ by gauge transformations; if $X=(x_1,\ldots,x_l) \in \curly G$ then the action is  
$$X\rho = ({\mathstrut x}_1{\mathstrut g}_{12}x_2^{\scriptscriptstyle{-1}},\ldots,{\mathstrut x}_i{\mathstrut g}_{ij}x_j^{\scriptscriptstyle{-1}}\ldots ).$$
In examples we have examined, this left action makes $\curly L$ into a fibre bundle, i.e., it is locally trivial, and we assume this is true, at least piecewise. Then the integration can be done in coordinates in which the bundle is trivialised. If several coordinate patches are required then the results can be glued together, for example using a partition of unity.

Let $\rho$ denote a fixed flat connection and set $\curly F=\curly G\rho$, the fibre of the bundle. For the proof it is sufficient to assume that $\curly L\cong \curly F\times \curly B$, where $\curly B$ is the base space, with maps $P\colon\curly L\to\curly B$ and $Q\colon\curly L\to\curly F$ which intertwine the actions of $\curly G$ (the action is trivial on $\curly B$).

Next, it is shown that the form $\lambda$ is invariant under the action of $X\in\curly G$. This follows from the facts that $\mu$ and $\nu$ both are. Acting with $X$ on (\ref{lambda}), one obtains
$$\mu = X^*\lambda \wedge F^*(\nu) \, .$$ 
Since (\ref{lambda}) determines $\lambda$ uniquely, then it follows that 
$$X^*\lambda=\lambda.$$

Let $\psi$ be a $\curly G$-invariant volume form on the fibre $\curly F$. Then this determines a unique form $\alpha$ given by the product of differentials of coordinates on $\curly B$ and functions on $\curly F\times \curly B$, by the equation
$$\alpha\wedge Q^*(\psi)=\lambda$$
 However a similar argument to that given above shows that $X^*\alpha=\alpha$ for any $X\in \curly G$, and so $\alpha$ is independent of the coordinates on $\curly F$. Thus it is the pullback of a volume form $\beta$ on $\curly B$, giving
\begin{equation}\label{lambdasplit}P^*(\beta)\wedge Q^*(\psi)=\lambda.\end{equation}
Integration then gives
\begin{equation}\label{measures}\int_{\curly L}|\lambda|=\int_{\curly B} |\beta|\int_{\curly F}|\psi|.\end{equation}

If $\curly F \cong \curly G$ then $\psi$ would be just the Haar measure on $\curly G$. However, in general, there is a stability subgroup $\curly K \subset \curly G$ with $k\rho = \rho$. So with fixed $\rho\in\curly F$ define a map of $\curly G$ to $\curly F$ by
$$A\colon X\mapsto X\rho.$$ 
Let $\dd k$ be the Haar measure on $\curly K$ and $\eta=\dd x_1\wedge\ldots\dd x_l$ the Haar measure on $\curly G$. The volume form $\dd k$ can be extended to a differential form $\widetilde{\dd k}$ on $\curly G$ which is left-invariant and agrees with $\dd k$ by pull-back to $\curly K$.

Then the form $\psi$ on $\curly F$ is defined by the formula
\begin{equation}\label{psi}\eta= A^*(\psi)\wedge \widetilde{\dd k}.\end{equation} 
Since $A$ commutes with the left action of $\curly G$, it follows that $\psi$ is also left-invariant. 

The map $A$ gives $\curly G$ a fibre bundle structure, with base space $\curly F$. Integrating along a fibre gives
$$\int_{X\curly K}\;\widetilde{\dd k}=\int_{\curly K}\;\dd k=1.$$
Hence
(\ref{psi}) gives 
$$1=\int_{\curly G}\eta=\int_{\curly F} \psi,$$
and (\ref{measures}) gives the desired integral as
$$\int_{\curly L}|\lambda|=\int_{\curly B} |\beta|.$$

Thus it only remains to determine a useful formula for the form $\beta$. Since it is independent of the point $X\in\curly G$, the calculation can be done at the identity, $X=I$.
It is necessary to relate a 0-chain $\Phi$ to a tangent vector to $\curly G$ by the isomorphism
$$\Xi_0\colon C^0(K)\to T\curly G_X$$
determined by its value on the $i$-th 0-cell $\sigma$,
$$\Phi(\sigma)=\frac{\dd {\mathstrut x}_i}{\dd t}.$$
Applying (\ref{psi}) gives
$$\tau_0\;<\dd x,\mathbf{l}>^l\;=\;<\Xi^*_0\,\eta,\mathbf{h}^0\mathbf{b}^0>\;
=\;<\Xi^*_0\,A^*\psi,\mathbf{b}^0>\,<\Xi^*_0\,\widetilde{\dd k},\mathbf{h}^0>.$$
A calculation shows that $A_*\Xi_0=-\Xi_1d_0$, which implies
$$<\Xi^*_0\,A^*\psi,\mathbf{b}^0>\;=\;<\Xi^*_1\,\psi,d_0\mathbf{b}^0>,$$
as the minus sign does not affect the $<\cdot,\cdot>$ evaluation.
From (\ref{lambdasplit}),
$$<\Xi^*_1\,\lambda,d_0(\mathbf{b}^0)\mathbf{\widetilde{h^1}}>\;=\;<\Xi^*_1\,\beta,\mathbf{h}^1><\Xi^*_1\,\psi,d_0\mathbf{b}^0>.$$
The first mapping $\Xi^*_1$ on the right-hand side refers to the induced mapping of the quotient $H^1(K)$ to the tangent space of $\curly B$.

Finally, $-(m-n-l)=\chi(K)=\chi(T(L))=\chi(T(\Gamma))=\chi(\Gamma)$, the Euler characteristic of the graph. Putting everything together leads to the formula
\begin{equation}\label{bakedcake}<\Xi^*_1\,\beta,\mathbf{h}^1>\;=\;\tor(T(L))\;<\Xi^*_0\,{\dd k},\mathbf{h}^0>\, <\dd g,\mathbf{l}>^{-\chi(\Gamma)}\;\prod_{e \in \Gamma} \delta\left(\theta_e - c(h_e)\right) .\end{equation}
This formula defines the differential form $\beta$. In fact, if $z_1,\ldots, z_n$ are coordinates on $\curly B$ and ${\Xi_{1}}_*\,\mathbf{h}^1$ is the basis $\frac{\partial}{\partial z_1},\ldots,\frac{\partial}{\partial z_n}$, then
$$\beta=\;<\beta,{\Xi_{1}}_*\,\mathbf{h}^1>\;\dd z_1\wedge\ldots\wedge\dd z_n.$$

The formula (\ref{bakedcake}) can be understood as follows. The torsion is a topological invariant.
 A vector in $C^0(K)$ lying in $H^0$ is invariant under the holonomy and so lies in the Lie algebra of the stability subgroup $\curly K$ of $\rho\in\curly G$. Thus $\mathbf{h}^0$ is a basis of $\mathrm{Lie}(\curly K)$. The factor $<\dd g,\mathbf{l}>$ is a numerical constant just depending on the choice of basis in the Lie algebra and corrects for the dependence of the torsion on the scaling of this basis.

The overall result can be summarised as follows.
\begin{theorem}
Suppose the parameters $\theta$ in the partition function (\ref{partition}) satisfy the existence criterion of section \ref{existence}, then the partition function is defined by an integral on the space $\curly B$ of flat connections modulo gauge transformations
\begin{equation}\label{Zistorsion}
Z=\int_{\curly B}|\beta|
\end{equation}
where the form $\beta$ is defined at a point $P(\rho)\in\curly B$ by (\ref{bakedcake}).
This partition function is independent of the choices of triangulation of the manifold and choices in the regularisation which are manifest in (\ref{partition}). It determines a topological invariant of the graph in the manifold.
\end{theorem}

\section{Calculation}\label{calculation}
\subsection{CW complexes}

A cell complex, such as the dual cell complex used above, is composed of subsets, the cells, each of which is isomorphic to a convex polyhedral ball. The key requirement is that if $\sigma_1$ and $\sigma_2$ are cells, then $\sigma_1\cap\sigma_2$ is a complete face in the boundary of both cells, of course of some lower dimension. 

In fact, computations are most easily carried out with the more general notion of a CW complex. A CW complex is constructed inductively, starting with a set of points (0-cells) and then attaching cells one by one using a map of their boundary onto the complex already constructed. The cells are attached in order of increasing dimension, so the 1-cells are all attached before the 2-cells, etc. The key generalisation is that the attaching map need not be 1-1, so that many points on the boundary of the new cell may be attached to the same point of the complex. Also, the intersection of two cells need no longer be a cell and the boundary of a cell need not be a union of cells (of lower dimension).

However in all the examples considered here, the boundary of a cell is always a union of cells. This makes it easy to formulate the twisted cohomology in the same way as for the simplicial case discussed in section \ref{twistedcohomology}, with the one proviso that the basepoint $v(\sigma)$ of cell $\sigma$ has to be chosen \emph{before} the cell is attached. This distinguishes the different vertices of the cell which may become identified by the attaching.

\subsubsection{Example - twisted cohomology of the circle} We calculate the twisted cohomology of the circle. The circle can be presented as a CW complex with one vertex $v$ and one 1-cell $f$. Choose an orientation for $f$. The flat connection is determined by the element $g_f\in\SU(2)$.  

First, consider the case $g_f\ne\pm \identity$.
Let the element of $\SO(3)$ corresponding to $g_f$ have axis of rotation the unit vector $\mathbf{n} \in \mathbb{R}^3$. Let $\Phi \in C^0$ defined by $\Phi(v) = \mathbf{x}.\mathbf{\sigma}$ where $\mathbf{x} \in \mathbb{R}^3$ and $\mathbf{\sigma}$ are the Pauli matrices, an orthogonal basis for the Lie algebra. We have
\begin{equation*}
\dd\Phi(f) = (g_f - \identity)\Phi(v).
\end{equation*}
This is equal to zero if and only if $\mathbf{x}$ is parallel to $\mathbf{n}$. So
\begin{equation*}
H^0 \; = \; \rm{Ker}\, \dd_0 \; \cong \; \mathbb{R}.
\end{equation*}
Also, $\rm{Im}\, \dd_0 = \lbrace \Psi \in C^1 \, \lvert \, \Psi(f) = \mathbf{y}.\mathbf{\sigma}, \: \mathbf{y} \in \mathbb{R}^3, \: \mathbf{y}.\mathbf{n} = 0 \rbrace$, and $\rm{Ker}\, \dd_1 = C^1$, so
\begin{equation*}
H^1 \; = \; C^1/\rm{Im}\, \dd_0 \; \cong \; \mathbb{R}.
\end{equation*}
The cohomology is different for the cases $g_f=\pm \identity$. In these cases, the twisted cohomology is the same as the ordinary cohomology, as $\pm \identity\in\SU(2)$ both act as the identity in the adjoint action. Hence
\begin{equation*}
H^0  \cong  \mathbb{R}^3 \quad\text{and}\quad H^1  \cong  \mathbb{R}^3.
\end{equation*}

\subsection{Examples - partition function by group integration} \label{calculationofZ}
In the following examples, we use a CW-complex $L$ for the complement ${S}^3\setminus \mathrm{interior} (T(\Gamma))$ determined by a diagram for the graph $\Gamma$. The information in this CW-complex can be reduced to a particular presentation of the fundamental group $\Pi_1({S}^3 \setminus \Gamma)$ which we call the region presentation. This is done by collapsing $L$ down to a 2-complex $K$, exactly as in section \ref{collapsing}.

\subsubsection{CW-complexes and group presentations}

From a CW-complex $K$, it is possible to read off a presentation of the fundamental group $\Pi_1(K)$. This is done by picking a vertex to be a basepoint and a maximal tree of edges. The generators of $\Pi_1(K)$ are then the remaining edges and the relations are given by the 2-cells, whose boundaries determine words in the generators. Any higher dimensional cells play no role in this process. Conversely, given a group presentation, then one can construct a 2-complex with one vertex, a loop for every generator and a 2-cell for every relation, attached along the corresponding sequence of edges. For 2-complexes, these two constructions are essentially inverse to each other (for more detail see \cite[section I-1]{cxbook}).

\subsubsection{The region presentation}

Let $B^k$ denote the $k$-dimensional ball. A $k$-handle is a 3-cell which is a thickened $k$-ball, $B^k\times B^{3-k}$, the second factor denoting the thickening up to three dimensions. The core of the handle is the subset $B^k\times\{p\}$, with $p$ an interior point of $B^{3-k}$. All 3-cells are of course topologically 3-balls but the distinction between them is the way in which they are attached. A handle decomposition of a manifold is a cellular complex which is constructed inductively by starting with a 0-handle and attaching handles one by one. A $k$-handle is attached by the thickened boundary of its core, $(\partial B^k)\times B^{3-k}$.

The construction of the region presentation is as follows.
First, a canonical handle decomposition of the manifold ${S}^3\setminus \mathrm{interior} (T(\Gamma))$ is constructed from the given diagram for the graph $\Gamma$, containing 0-,1- and 2-handles. Then this manifold can be collapsed to a 2-complex consisting of the cores of all of the handles. This 2-complex is determined by a corresponding presentation of $\Pi_1$. First the resulting presentation is described, then the argument leading to it is given.

\begin{figure}[h]
$$\epsfbox{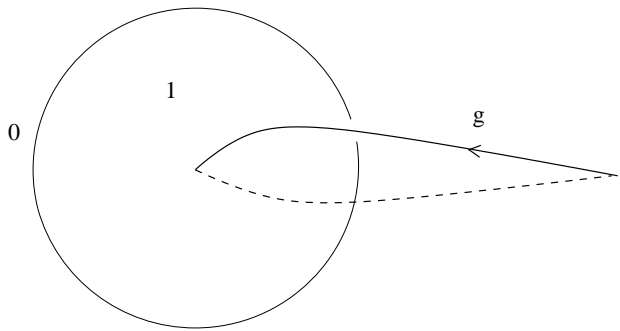}$$
 \caption{The fundamental group of the unknot complement has a single generator.}
 \label{unknotgenerator}
\end{figure}

\begin{figure}[h]
$$\epsfbox{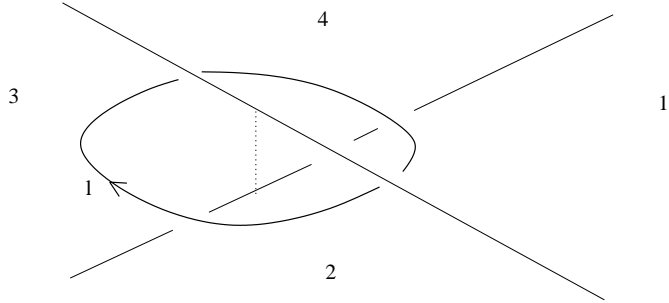}$$
 \caption{A typical crossing. Also depicted is an extra edge joining the over- and under-lying edges at the crossing. Equating $l$ with the identity loop gives $g_1 \, g_2^{-1} \, g_3 \,g_4^{-1} = \identity$.}
 \label{crossingrelation}
\end{figure}

\begin{figure}[h]
$$\epsfbox{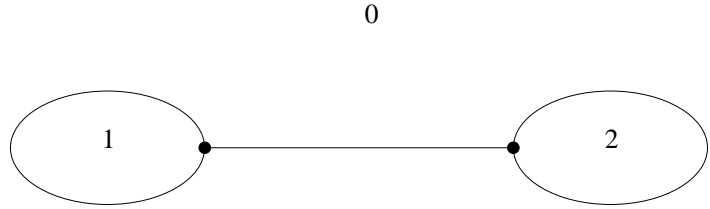}$$
 \caption{The dumbell graph}
\label{dumbellgraph}
\end{figure}

The `region' presentation of $\Pi_1({S}^3 \setminus \Gamma)$ is defined as follows. Begin by numbering the regions of the graph, assigning 0 to the outer region (see for example figure \ref{dumbellgraph}). Then there is one generator for each region of the graph, except the outer region, determined by the loop that starts above the graph, pierces that region, and then returns to the start point by piercing the outer region. This is illustrated for the unknot in figure \ref{unknotgenerator}. There is one relation for each crossing point (so planar graphs have no relations): at each crossing, imagine inserting an extra edge to join the upper and lower parts. The element of $\Pi_1$ corresponding to the loop encircling that edge, in a clockwise direction as viewed from above, is equivalent to a product of generators corresponding to the regions which meet at the crossing point. The expression equating this loop with the identity loop gives the relation for the crossing point (see figure \ref{crossingrelation}). 

The CW-complex corresponding to this presentation of $\Pi_1({S}^3 \setminus \Gamma)$ has one 0-cell, a 1-cell for each generator, and a 2-cell for each relation, attached along the relation. The orientations of the cells are given by the loops defining them.  We write $h_i$ for the holonomy along the 1-cell corresponding to the $i$th region.

The argument that this construction is correct is as follows. Consider first a planar graph $\Gamma$ and a diagram of it without crossings. This can be considered a graph on $S^2$ and the complement of the thickening $T(\Gamma)$ has a decomposition as two zero-handles (one above the diagram and one below it) and a 1-handle for each region of $S^2 \setminus \Gamma$. Each 1-handle is attached to both 0-handles.

Now consider the general case of a graph $\Gamma$ with crossings. On the diagram, each crossing point can be replaced with a 4-valent vertex to give a planar graph $\Gamma'$. The thickening $T(\Gamma')$ is homeomorphic to a thickening of $\Gamma$ with an extra edge added at each crossing, as in figure \ref{crossingrelation}. Therefore a handle decomposition of ${S}^3\setminus \mathrm{interior} (T(\Gamma))$ can be constructed from the handle decomposition of the planar ${S}^3\setminus \mathrm{interior} (T(\Gamma'))$ by adding a 2-handle for each of the extra edges at the crossings. This is because a handle decomposition for $T(\Gamma')$ and its complement together determine a handle decomposition for $S^3$, and a 1-handle for $T(\Gamma')$ can alternatively be viewed as a 2-handle for 
${S}^3\setminus \mathrm{interior} (T(\Gamma))$.

This constructs a handle decomposition for ${S}^3\setminus \mathrm{interior} (T(\Gamma))$ in which there are two 0-handles, a 1-handle for every region and a 2-handle for every crossing. The two 0-handles and the 1-handle for the outside region (labelled 0) can be merged to form a single 0-handle. Now collapsing every handle to its core yields a 2-complex with one vertex, and its associated presentation is the region presentation of $\Pi_1({S}^3 \setminus \Gamma)$ described above.

\subsubsection{Planar graphs}

For the region presentation of the graph complement, the are no 2-cells if the graph is a planar graph. Therefore for a planar graph the twisted $H^2$ is always zero and the partition function always exists.

\subsubsection{Dumbell graph}
The CW-complex for the dumbell graph has two 1-cells corresponding to regions 1 and 2. The formula for $Z$ gives
\begin{align*}
Z &= \int \dd h_{1} \, \dd h_{2} \: \delta(\theta_{00}) \delta \left( c(h_{1}) - \theta_{01} \right) \delta \left( c(h_{2}) - \theta_{02} \right) \\
&= \bigl( \tfrac{1}{\pi} \bigr)^2 \sin^{2}\left(\tfrac{1}{2}\theta_{01}\right) \sin^{2}\left(\tfrac{1}{2}\theta_{02}\right) \: \delta(\theta_{00}).
\end{align*}

\subsubsection{Theta graph}
\begin{figure}[h]
$$\epsfbox{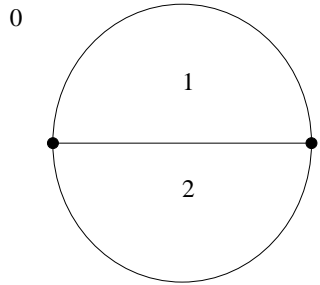}$$
 \caption{The theta graph}
\end{figure}
To calculate the theta graph we follow a procedure similar to that employed by Freidel and Louapre in their calculation of the tetrahedron (see \cite{FLA}). We have
\begin{equation}\label{theta1}
Z = \int \dd h_{1} \, \dd h_{2} \: \delta \left( c(h_{1}) - \theta_{01} \right) \delta \left( c(h_{2}) - \theta_{02} \right) \delta \left( c(h_{1}h_{2}^{\scriptstyle{-1}}) - \theta_{12} \right).
\end{equation}
The integrand is invariant under the transformations
\begin{equation*}
h_{i} \quad \rightarrow \quad k \, h_{i} \, k^{\scriptscriptstyle{-1}},
\end{equation*} 
where $k \in \SU(2)$. We will gauge out this symmetry and write the integral in terms of the gauge invariant variables $c(h_{1}), c(h_{2}), c(h_{1}h_{2}^{\scriptscriptstyle{-1}})$. The gauge fixing will be described in relation to the geometrical situation: the two group variables $h_{1}, h_{2}$, define two points in ${S}^3$ and these together with the origin define a triangle in ${S}^{3}$. The invariant geometry of this triangle is parametrised by its edge lengths $\phi_{1} = c(h_{1}), \, \phi_{2} = c(h_{2}), \,  \phi_{12} = c(h_{1}h_{2}^{\scriptscriptstyle{-1}})$. We will also need the angle $\tilde\phi_{12} \in [0,\pi]$ defined by $\cos(\tilde\phi_{12}) = \mathbf{n}_{1}.\mathbf{n}_{2}$, where $\mathbf{n}_{i}$ is the axis of rotation of the element of $\SO(3)$ corresponding to $h_{i}$. This obeys the relation
\begin{equation}\label{phi12}
\cos\left(\tfrac{1}{2}\phi_{12}\right) = \cos\left(\tfrac{1}{2}\phi_{1}\right)\cos\left(\tfrac{1}{2}\phi_{2}\right) + \sin\left(\tfrac{1}{2}\phi_{1}\right)\sin\left(\tfrac{1}{2}\phi_{2}\right)\cos\bigl(\tilde\phi_{12}\bigr),
\end{equation}
(which is readily verified by calculating $\cos\bigl(\tfrac{1}{2}\phi_{12}\bigr) = \tfrac{1}{2}\mathrm{Tr}\bigl(h_{1}h_{2}^{\scriptscriptstyle{-1}}\bigr)$). The gauge fixing is in two steps. First we rotate the triangle about the origin so that $\mathbf{n}_{1}$ lies along the $x$-axis. Second we rotate about $\mathbf{n}_{1}$ so that the triangle lies in the $xy$-plane. That is, we fix 
\begin{equation}\label{gaugefix}
\mathbf{n}_{1} = 
\begin{pmatrix} 
1 \\ 0 \\ 0 
\end{pmatrix}
, \quad \quad \mathbf{n}_{2} =
\begin{pmatrix} 
\cos\bigl(\tilde\phi_{12}\bigr) \\ \sin\bigl(\tilde\phi_{12}\bigr) \\ 0 
\end{pmatrix}.
\end{equation}
Now if $(\phi, \mathbf{n})$ are the angle and axis of rotation for the element of \SO(3) corresponding to $g\in\SU(2)$ then
\begin{equation*}
\dd g = \frac{1}{\pi} \: \sin^{2}\left(\tfrac{1}{2}\phi\right) \: \dd\phi \: \dd\mathbf{n},
\end{equation*}
where $\dd\mathbf{n}$ is the invariant measure on ${S}^2$ with $\smallint \dd\mathbf{n} = 1$. But from (\ref{phi12}) we have
\begin{equation*}
\tfrac{1}{2}\sin\left(\tfrac{1}{2}\phi_{12}\right) \dd\phi_{12} = \sin\left(\tfrac{1}{2}\phi_{1}\right)\sin\left(\tfrac{1}{2}\phi_{2}\right)\sin\bigl(\tilde\phi_{12}\bigr) \dd\tilde\phi_{12},
\end{equation*}
so in terms of the gauge invariant observables, the measure is
\begin{equation*}
\dd h_{1} \, \dd h_{2} = \bigl( \tfrac{1}{\pi} \bigr)^2  \sin\left(\tfrac{1}{2}\phi_{1} \right) \sin\left(\tfrac{1}{2}\phi_{2} \right) \dd\phi_{1} \dd\phi_{2} \;\tfrac{1}{4} \sin\left(\tfrac{1}{2}\phi_{12}\right) \dd\phi_{12}.
\end{equation*}
Substituting in (\ref{theta1}) leads immediately to 
\begin{equation*}
Z = \begin{cases}
	\left(\tfrac{1}{2\pi}\right)^2 \sin\left(\tfrac{1}{2}\theta_{01} \right) \sin\left(\tfrac{1}{2}	\theta_{02} \right) \sin\left(\tfrac{1}{2}\theta_{12}\right) & \text{if $\lbrace\theta_{ij}\vert i<j\rbrace$ satisfy triangle inequalities.}
	\\ 0 & \text{otherwise.}
	\end{cases}	
\end{equation*}

\subsubsection{Humbug graph}
\begin{figure}[h]
$$\epsfbox{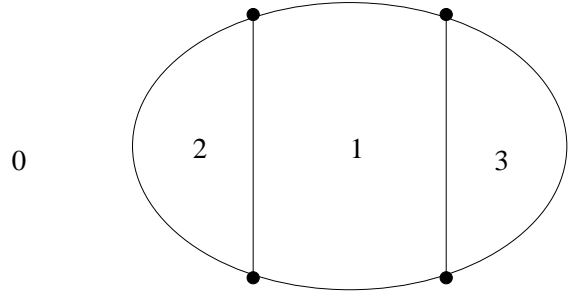}$$
 \caption{The humbug graph}
 \label{humbug}
\end{figure}
Referring to figure \ref{humbug}, variables $\theta_{01}$ and $\theta'_{01}$ label the two edges which border regions 0 and 1. We have
\begin{equation}\label{humbug1} \begin{split}
Z = \int &\dd h_{1} \, \dd h_{2} \, \dd h_{3} \: \delta\bigl( c(h_{1}) - \theta_{01} \bigr) \delta\bigl( c(h_{1}) - \theta_{01}' \bigr) \delta\bigl( c(h_{2}) - \theta_{02} \bigr) \\ 
&\delta\bigl( c(h_{3}) - \theta_{03} \bigr) \delta\bigl( c(h_{2}h_{1}^{-1}) - \theta_{12} \bigr) \delta\bigl( c(h_{3}h_{1}^{-1}) - \theta_{13} \bigr)  \end{split}
\end{equation}
This may be evaluated using a similar gauge fixing procedure to that for the theta graph. Define new angles $\tilde\phi_{12}$, $\tilde\phi_{13} \in [0,\pi]$ by $\cos(\tilde\phi_{12}) = \mathbf{n}_{1}.\mathbf{n}_{2}, \, \cos(\tilde\phi_{13}) = \mathbf{n}_{1}.\mathbf{n}_{3}$. Then (\ref{humbug1}) becomes
\begin{align*} 
Z = \int &\left(\tfrac{1}{\pi}\right)^3 \sin^2\left(\tfrac{1}{2}\phi_{1} \right) \sin^2\left(\tfrac{1}{2}\phi_{2} \right) \sin^2\left(\tfrac{1}{2}\phi_{3} \right) \dd\phi_{1} \, \dd\phi_{2} \, \dd\phi_{3} \tfrac{1}{4} \sin\bigl(\tilde\phi_{12}\bigr) \sin\bigl(\tilde\phi_{13}\bigr) \dd\tilde\phi_{12} \, \dd\tilde\phi_{13} \\ 
&\delta\left( \phi_{1} - \theta_{01} \right) \delta\left( \phi_{1} - \theta_{01}' \right) \delta\left( \phi_{2} - \theta_{02} \right) \delta\left( \phi_{3} - \theta_{03} \right) \delta\left(\phi_{12} - \theta_{12} \right) \delta\left( \phi_{13} - \theta_{13} \right),
\end{align*}
which, using the relations analogous to (\ref{phi12}), evaluates to
\begin{equation*}
Z = \left\{ \begin{array}{ll}
\tfrac{1}{2} \left(\tfrac{1}{2\pi}\right)^3 \sin\left(\tfrac{1}{2}\theta_{03} \right) \sin\left(\tfrac{1}{2} \theta_{02} \right) &\;\text{if $\left(\theta_{01},\theta_{02},\theta_{12}\right)$, $\left(\theta_{01},\theta_{03},\theta_{13}\right)$ satisfy}\\
\quad\sin\left(\tfrac{1}{2}\theta_{12}\right)\sin\left(\tfrac{1}{2}\theta_{13}\right) \delta\left( \theta_{01} - \theta_{01}' \right) &\quad\;\text{triangle inequalities.}\\
0 &\;\text{otherwise.}	
\end{array} \right.
\end{equation*}

\subsubsection{Unknot}
\begin{figure}[h]
$$\epsfbox{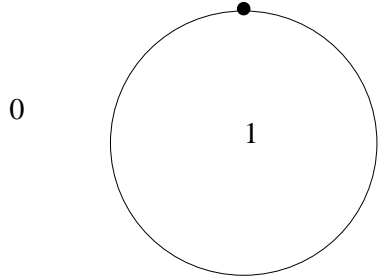}$$
 \caption{The unknot}
 \label{unknot}
\end{figure}
Referring to figure \ref{unknot}, we have
\begin{equation}
Z \; = \; \int \dd h_{1} \: \delta \bigl( c(h_{1}) - \theta_{01} \bigr) \; = \; \tfrac{1}{\pi} \, \sin^{2}\left(\tfrac{1}{2}\theta_{01} \right).
\end{equation}

\subsubsection{Trefoil knot}
\begin{figure}[h]
$$\epsfbox{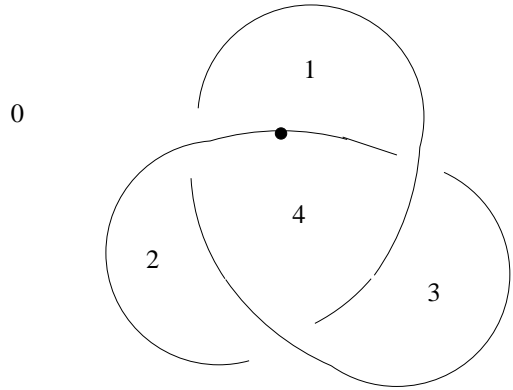}$$
 \caption{The trefoil knot}
 \label{trefoil}
\end{figure}
Referring to figure \ref{trefoil}, we have
\begin{equation*}
Z = \int \dd h_{1} \, \dd h_{2} \, \dd h_{3} \, \dd h_{4} \; \delta\bigl( c(h_{1}) - \theta \bigr) \delta\left( h_{2}h_{4}^{\scriptscriptstyle{-1}}h_{1} \right) \delta\left( h_{3}^{\scriptscriptstyle{-1}}h_{4}h_{1}^{\scriptscriptstyle{-1}} \right) \delta\left( h_{3}h_{4}^{\scriptscriptstyle{-1}}h_{2} \right). 
\end{equation*}
Here there is only one delta-function of the holonomy fixing type since the knot has just one vertex. Eliminate $h_{4}$ using the final delta-function to get
\begin{equation*}
Z = \int \dd h_{1} \, \dd h_{2} \, \dd h_{3} \; \delta\bigl( c(h_{1}) - \theta \bigr) \delta\left( h_{2}h_{3}^{\scriptscriptstyle{-1}}h_{2}^{\scriptscriptstyle{-1}}h_{1} \right) \delta\left( h_{3}^{\scriptscriptstyle{-1}}h_{2}h_{3}h_{1}^{\scriptscriptstyle{-1}} \right). 
\end{equation*}
The flat connections on the trefoil exterior (ie. the solutions to the relations imposed by the delta-functions) split into two branches. The abelian branch $\rho_A = \lbrace h_{1}, h_{2}, h_{3} \vert h_{1} = h_{2} = h_{3} \rbrace$ exists for all $\theta \in [0, 2\pi]$ and has $H^2(L,\rho_A) = 0$, whilst the non-abelian branch $\rho_{NA}$ exists for $\theta \in [\pi/3,5\pi/3]$ and has $H^2(L,\rho_{NA}) = \R$ (see \cite{DUB}). So $Z$ exists only for $\theta <   \pi/3$, $\theta > 5\pi/3$, and in this range, by linearising around the solution $h_{1} = h_{2} = h_{3}$ one obtains \cite{B2}
\begin{equation}\label{Ztrefoil}
Z = \frac{1}{\pi} \sin^2\left(\tfrac{1}{2}\theta\right) \frac{1}{\lvert D \rvert} \quad\quad\quad \theta <   \pi/3, \; \theta > 5\pi/3
\end{equation}
where $D$ is the 6 $\times$ 6 determinant
\begin{equation*}
D = \begin{vmatrix}
	1-X &X\\
	-1 &1-X
	\end{vmatrix}
= |1-X+X^2|,
\end{equation*}	
where $X$ is the 3 $\times$ 3 matrix for the element of $\SO(3)$ corresponding to $h_{1}$. The polynomial $1-X+X^2$ is the Alexander polynomial of the trefoil knot. Since $X$ has eigenvalues $e^{i\theta}$, $e^{-i\theta}$ and $1$, the determinant is
$$|1-X+X^2|=|1-e^{i\theta}+e^{2i\theta}|^2.$$

\subsubsection{Figure-eight knot}
\begin{figure}[h]
$$\epsfbox{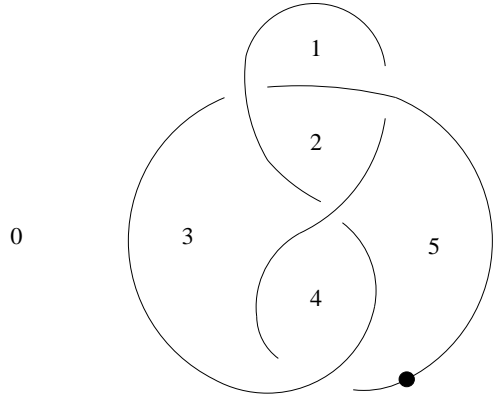}$$
 \caption{The figure-eight knot}
\label{figure-eight}
\end{figure}
Referring to figure \ref{figure-eight} we have
\begin{align*}
Z = \int &\dd h_{1} \ldots \dd h_{5}\; \delta\bigl( c(h_{1}) - \theta \bigr) \delta\left( h_{4}h_{3}^{\scriptscriptstyle{-1}}h_{2}h_{5}^{\scriptscriptstyle{-1}} \right) \delta\left( h_{3}^{\scriptscriptstyle{-1}}h_{4}h_{5}^{\scriptscriptstyle{-1}} \right)\\
& \delta\left(h_{5}^{\scriptscriptstyle{-1}}h_{2}h_{1}^{\scriptscriptstyle{-1}}  \right)\delta\left(h_{1}^{\scriptscriptstyle{-1}}h_{2}h_{3}^{\scriptscriptstyle{-1}}  \right)
\end{align*}
Eliminate $h_{3}$ by integrating out the final delta function and then make the change of variables $h_{2} \rightarrow h_{52} = h_{2}h_{5}^{\scriptscriptstyle{-1}}$, $h_{4} \rightarrow h_{45} = h_{5}h_{4}^{\scriptscriptstyle{-1}}$, followed by the change of variable $h_{5} \rightarrow h_{5}^{\scriptscriptstyle{-1}}$ to get
\begin{align*}
Z = \int &\dd h_{1} \, \dd h_{52} \,  \dd h_{45} \,  \dd h_{5} \; \delta\bigl( c(h_{1}) - \theta \bigr) \delta\left( h_{45}^{\scriptscriptstyle{-1}}h_{52}^{\scriptscriptstyle{-1}}h_{1}h_{52} \right) \\
&\delta\left( h_{5}h_{45}h_{52}^{\scriptscriptstyle{-1}}h_{45}^{\scriptscriptstyle{-1}} \right) \delta\left(h_{5}h_{52}h_{5}^{\scriptscriptstyle{-1}}h_{1}^{\scriptscriptstyle{-1}}  \right)
\end{align*}
As with the trefoil knot, the flat connections on the figure-eight exterior split into an abelian branch $\rho_{A}$ and a non-abelian branch $\rho_{NA}$ for which $H^2(L,\rho_A) = 0$ and $H^2(L,\rho_{NA}) = \R$ (see \cite{DUB}). The abelian branch exists for all $\theta \in [0, 2\pi]$, the non-abelian branch for $\theta \in [2\pi/5,8\pi/5]$. So $Z$ exists only for $\theta <   2\pi/5$, $\theta > 8\pi/5$, and in this range, linearising around the abelian solution gives
\begin{equation}\label{Zfigure-eight}
Z = \frac{1}{\pi} \sin^2\left(\tfrac{1}{2}\theta\right) \frac{1}{\lvert D \rvert} \quad\quad\quad \theta <   2\pi/5, \; \theta > 8\pi/5
\end{equation}
where $D$ is the 9 $\times$ 9 determinant
\begin{equation*}
D = \begin{vmatrix}
	X &1-X &-1\\
	X &0 &1-X\\
	1-X &X &0
	\end{vmatrix}
=|1-3X+X^2|,
\end{equation*}	
and where $X$ is the 3 $\times$ 3 matrix for the element of $\SO(3)$ corresponding to $h_{1}$. Again, the polynomial in this calculation, $1-3X+X^2$, is the Alexander polynomial of the knot.

\subsection{Examples - the Reidemeister torsion for the trefoil and figure-eight knots}\label{calculationoftor}

The partition function can be calculated using the Reidemeister torsion by specialising formulae (\ref{Zistorsion}),(\ref{bakedcake}) to the case of a knot. Since this is an invariant, it is possible to use any convenient CW complex for its calculation. The explicit calculations are done for two knots considered in the previous section but using different CW complexes. These complexes have the advantage that they generalise easily to any knot. The generalisation is carried out in section \ref{knotsandAlexpoly}.

Denoting our knot by $K$, we may calculate $\mathrm{tor}\left( L \right)$ using the CW-complex arising from the Wirtinger presentation of $\Pi_1({S}^3\setminus K)$ described as follows. Choose a knot diagram and a point $p$ above the diagram. Then for each arc of the diagram there is an element $x_i$ of $\Pi_1$ corresponding to the loop, with base point $p$, which encircles the arc once in the sense of a right-handed screw. The $x_i$ are the generators of the presentation. At each crossing there is a relation $r_j$. The relations are not all independent; any one may be derived from the others. Deleting one relation, one arrives at the Wirtinger presentation $\lbrace x_1, \ldots , x_n | r_1, \ldots , r_{n-1} \rbrace$. The corresponding CW-complex, $\mathcal{C}$, has one 0-cell, $V$, $n$ 1-cells, $X_i$ and $n-1$ 2-cells, $R_j$. The justification for the use of $\mathcal{C}$ is that one can construct a simple homotopy equivalence from the region presentation to $\mathcal{C}$. This is done in appendix 2. This is non-trivial because it is not true that any presentation of a group is simple homotopy equivalent to any other.

\subsubsection{Trefoil knot}

\begin{figure}[h]
$$\epsfbox{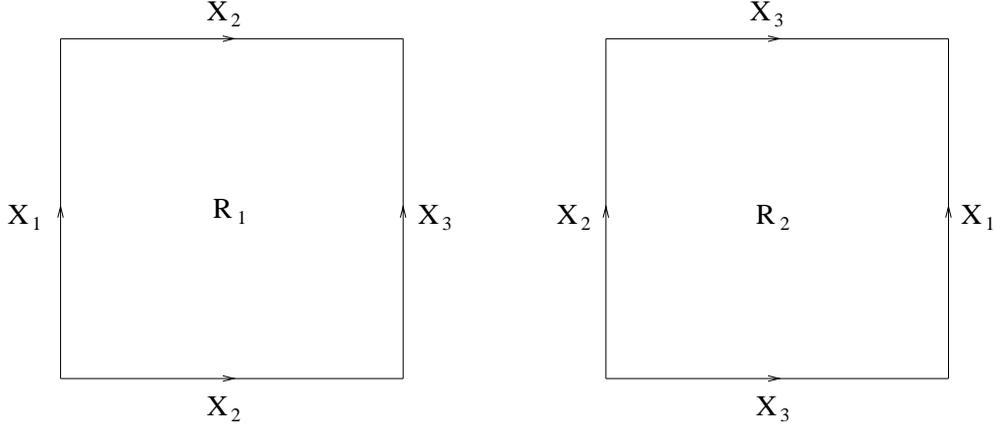}$$
 \caption{The 2-cells $R_i$ for the CW-complex of the trefoil exterior, attached along the relations $r_i$.}
 \label{trefoiltwocells}
\end{figure} 

Let $K$ denote the trefoil knot. The CW-complex for ${S}^3\setminus \mathrm{interior} (T(K))$ has one 0-cell, three 1-cells and two 2-cells, attached as shown in figure \ref{trefoiltwocells}. Orient the 1-cells so that their distinguished vertices are at their `from' ends. Choose the bottom left hand vertex of each 2-cell to be the distinguished vertex and orient the 2-cells so that their boundaries are oriented in an anti-clockwise direction.

Let $\rho$ be the abelian representation of $\Pi_1({S}^3\setminus K)$ with 
\begin{equation*}
\rho(x_i) = x, \quad i = 1,2,3, \quad\quad
x = \begin{pmatrix}
	e^{i\theta/2} & 0\\
	0 & e^{-i\theta/2}
	\end{pmatrix}.
\end{equation*}	
The basis $\mathbf l$ used for the Lie algebra is the set of Pauli matrices $\sigma_1, \sigma_2,\sigma_3$, with
$$\sigma_1=\begin{pmatrix}\frac12&0\\0&-\frac12\end{pmatrix},\quad
\sigma_2=\begin{pmatrix}0&\frac12\\\frac12&0\end{pmatrix},\quad
\sigma_3=\begin{pmatrix}0&-\frac i2\\\frac i2&0\end{pmatrix}.$$ 

Using the notation introduced in section \ref{Reidemeistertorsion}, define 
\begin{align*}
&\mathbf{c}^0 = \lbrace \Phi_1,\Phi_2,\Phi_3 \rbrace, \; \Phi_i(V) = \sigma_i \\
&\mathbf{c}^1 = \lbrace \Psi_{11}, \Psi_{21}, \Psi_{31}, \Psi_{12}, \Psi_{22}, \Psi_{32}, \Psi_{13}, \Psi_{23}, \Psi_{33} \rbrace, \; \Psi_{ij}(X_k) = \sigma_i \delta_{jk}\\
&\mathbf{c}^2 = \lbrace \Omega_{11},  \Omega_{21}, \Omega_{31}, \Omega_{12}, \Omega_{22}, \Omega_{32} \rbrace, \; \Omega_{ij}(R_k) = \sigma_i \delta_{jk}.
\end{align*} 
Then
\begin{align*}
&\dd_1\Psi_{i1} = -\, \Omega_{i1} + x\, \Omega_{i2}\\
&\dd_1\Psi_{i2} = (1-x)\, \Omega_{i1} - \Omega_{i2}\\
&\dd_1\Psi_{i3} = x\, \Omega_{i1} + (1-x)\, \Omega_{i2},\\
\intertext{and}
&\dd_0\Phi_i = (x-1) \sum_{j=1}^3 \Psi_{ij}.
\end{align*}
Choose $\mathbf{b}^1 = \lbrace \Psi_{12}, \Psi_{22}, \Psi_{32}, \Psi_{13}, \Psi_{23}, \Psi_{33} \rbrace, \: \mathbf{b}^0 = \lbrace \Phi_2, \Phi_3 \rbrace, \: \tilde{\mathbf{h}}^1 = \Psi_{11} + \Psi_{12} + \Psi_{13}, \: \tilde{\mathbf{h}}^0 = \Phi_1$. The factor in the Reidemeister torsion corresponding to $k=2$ in formula (\ref{tauk}) is
\begin{equation*}
\tau_2 = \left[ \dd_1(\mathbf{b}^1) / \mathbf{c}^2 \right]^{-1} =  \left[ \mathbf{a}^2 / \mathbf{c}^2 \right]^{-1}.
\end{equation*}
Now, $a^2_i = M^{(2)}_{ji}\,c^2_j$, where $M^{(2)}$ is the 6$\times$6 matrix
\begin{equation*}
M^{(2)} = \begin{pmatrix}
	1-X & X\\
	-1 & 1-X
	\end{pmatrix}
\end{equation*}
and where $X$ is the 3 $\times$ 3 matrix for the element of $\SO(3)$ corresponding to $x$. So $\tau_2 = \lvert M^{(2)} \rvert^{-1}$. The $k=1$ factor is
\begin{equation*}
\tau_1 = \left[ \dd_0(\mathbf{b}^0), \tilde{\mathbf{h}}^1,\mathbf{b}^1 / \mathbf{c}^1 \right] =  \left[ \mathbf{a}^1 / \mathbf{c}^1 \right].
\end{equation*}
Now, $a^1_i = M^{(1)}_{ij}\,c^1_j$, where $M^{(1)}$ is the 9$\times$9 matrix
\begin{equation*}
M^{(1)} = \begin{pmatrix}
	R & R & R\\
	0 & 1 & 0\\
	0 & 0 & 1
	\end{pmatrix},
\end{equation*}
and where $R$ is the 3$\times$3 matrix
\begin{equation}\label{R}
R = \begin{pmatrix}
	0 & \cos\theta -1 & -\sin\theta\\
	0 & \sin\theta & \cos\theta -1\\
	1 & 0 & 0
	\end{pmatrix}.
\end{equation}
So $\tau_1 = \lvert M^{(1)} \rvert = 4 \sin^2\left(\tfrac{1}{2}\theta\right)$. The factor for $k=0$ is
\begin{equation*}
\tau_0 = \left[ \tilde{\mathbf{h}}^0,\mathbf{b}^0 / \mathbf{c}^0 \right]^{-1} =  1.
\end{equation*}
Finally, multiplying the factors together gives
\begin{equation}\label{tortrefoil}
\mathrm{tor}(T(L)) = \frac{4\sin^2\left(\tfrac{1}{2}\theta\right)}{\begin{vmatrix}
	1-X & X\\
	-1 & 1-X
	\end{vmatrix}},
\end{equation}
Then plugging equation (\ref{tortrefoil}) into the formula for the partition function in terms of Reidemeister torsion (\ref{Zistorsion}) reproduces the result of our earlier calculation of the partition function for the trefoil knot (\ref{Ztrefoil}). Specifically, using $\theta$ as the fixed label in the partition function and $\theta'$ as the coordinate on $\curly B$, then
$$\chi(\Gamma)=0$$
$$\dd k=\frac1{4\pi}\dd\theta'$$
$$h^0=h^1=\frac\partial{\partial\theta'}$$
$$<\dd k,h^0>=\frac1{4\pi}$$
$$\beta=\frac1{4\pi}\mathrm{tor}(T(L))\delta(\theta-\theta')\dd\theta'$$
which gives
$$Z=\int_{\curly B}\beta=\frac1{4\pi}\mathrm{tor}(T(L))=\frac1\pi\frac{\sin^2\left(\tfrac{1}{2}\theta\right)}{\begin{vmatrix}
	1-X & X\\
	-1 & 1-X
	\end{vmatrix}},$$
as in section \ref{calculationofZ}.
\subsubsection{Figure-eight knot}

\begin{figure}[h]
$$\epsfbox{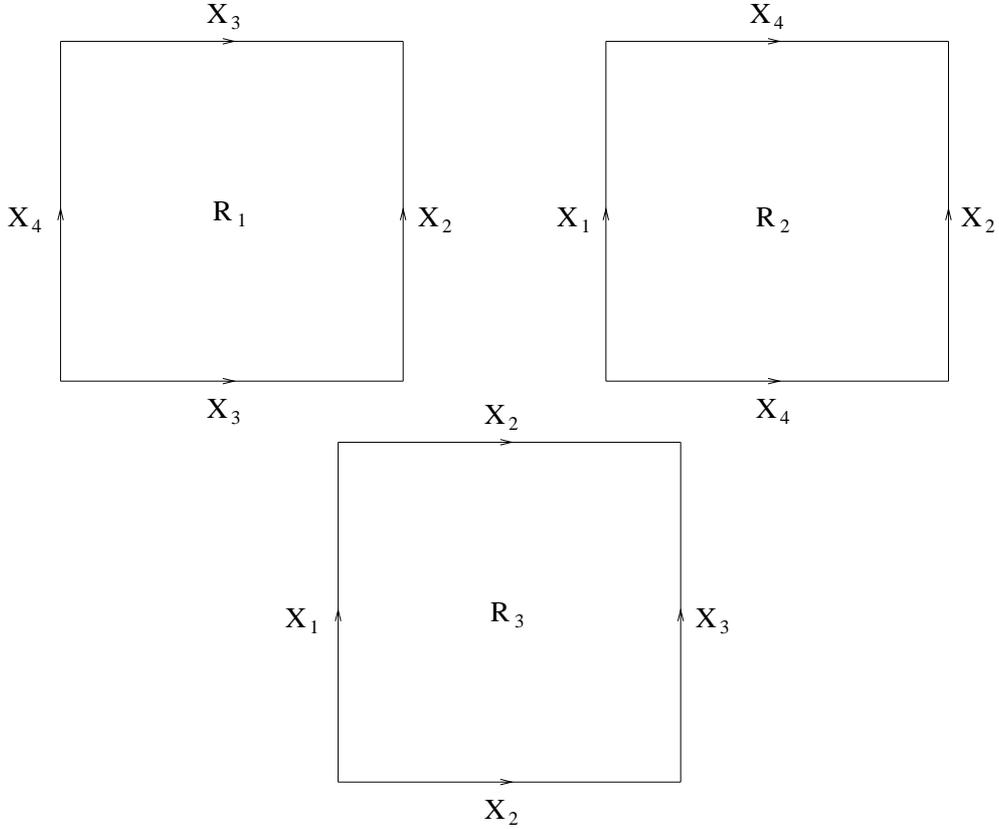}$$
 \caption{The 2-cells $R_i$ for the CW-complex of the figure-eight exterior, attached along the relations  $r_i$.}
 \label{figure-eighttwocells}
\end{figure} 

Let $K$ denote the figure-eight knot. The CW-complex for $ {S}^3\setminus \mathrm{interior} (T(K))$ has one 0-cell, four 1-cells and three 2-cells, attached as shown in figure \ref{figure-eighttwocells}. Let $\rho$ be the abelian representation of $\Pi_1({S}^3\setminus K)$ with 
\begin{equation*}
\rho(x_i) = x, \quad i = 1,2,3,4, \quad\quad
x = \begin{pmatrix}
	e^{i\theta/2} & 0\\
	0 & e^{-i\theta/2}
	\end{pmatrix}.
\end{equation*}	
Define 
\begin{align*}
&\mathbf{c}^0 = \lbrace \Phi_1,\Phi_2,\Phi_3 \rbrace, \; \Phi_i(V) = \sigma_i \\
&\mathbf{c}^1 = \lbrace \Psi_{11}, \Psi_{21}, \Psi_{31}, \ldots , \Psi_{14}, \Psi_{24}, \Psi_{34} \rbrace, \; \Psi_{ij}(X_k) = \sigma_i \delta_{jk}\\
&\mathbf{c}^2 = \lbrace \Omega_{11},  \Omega_{21}, \Omega_{31}, \ldots , \Omega_{13}, \Omega_{23}, \Omega_{33} \rbrace, \; \Omega_{ij}(R_k) = \sigma_i \delta_{jk}.
\end{align*} 
Calculating the matrices for $\dd_1$ and $\dd_0$ with respect to these bases suggests the following choices for the bases $\mathbf{a}^i$. 
\begin{gather*}
\mathbf{b}^1 = \lbrace \Psi_{12}, \Psi_{22}, \Psi_{32}, \Psi_{13}, \Psi_{23}, \Psi_{33}, \Psi_{14}, \Psi_{24}, \Psi_{34} \rbrace, \\ 
\tilde{\mathbf{h}}^1 = \Psi_{11} + \Psi_{12} + \Psi_{13} + \Psi_{14}, \: \mathbf{b}^0 = \lbrace \Phi_2, \Phi_3 \rbrace,  \: \tilde{\mathbf{h}}^0 = \Phi_1.
\end{gather*}
The factor in the Reidemeister torsion corresponding to $k=2$ in the formula (\ref{tauk}) is
\begin{equation*}
\tau_2 = \lvert M^{(2)} \rvert^{-1},
\end{equation*}
where $M^{(2)}$ is the 9$\times$9 matrix
\begin{equation*}
M^{(2)} = \begin{pmatrix}
	X & 1-X & -1\\
	X & 0 & 1-X\\
	1-X & X & 0
	\end{pmatrix}
\end{equation*}
and where $X$ is the 3 $\times$ 3 matrix for the element of $\SO(3)$ corresponding to $x$. The $k=1$ factor is
\begin{equation*}
\tau_1 =  \lvert M^{(1)} \rvert,
\end{equation*}
where $M^{(1)}$ is the 12$\times$12 matrix
\begin{equation*}
M^{(1)} = \begin{pmatrix}
	R & R & R & R\\
	0 & 1 & 0 & 0\\
	0 & 0 & 1 & 0\\
	0 & 0 & 0 & 1
	\end{pmatrix},
\end{equation*}
and where $R$ is the 3$\times$3 matrix defined in (\ref{R}). So $\tau_1 = 4 \sin^2\left(\tfrac{1}{2}\theta\right)$. As for the trefoil knot, the factor for $k=0$ is $\tau_0 = 1$. Finally, the Reidemeister torsion for the figure-eight exterior is given by
\begin{equation}\label{torfigure-eight}
\rm{tor} = \frac{4\sin^2\left(\tfrac{1}{2}\theta\right)}{\begin{vmatrix}
	X & 1-X & -1\\
	X & 0 & 1-X\\
	1-X & X & 0
	\end{vmatrix}}.
\end{equation}
Again, one may check that plugging this equation into formula (\ref{Zistorsion}) reproduces the result (\ref{Zfigure-eight}). 

\subsection{Knots and the Alexander polynomial} \label{knotsandAlexpoly}
We have seen in sections \ref{calculationofZ} and \ref{calculationoftor} that for $K$ either the trefoil knot or the figure-eight knot, $Z({S}^3, K)$ is calculated using the Alexander polyomial of $K$. In fact, the same result is true for any knot $K$ and follows from the specialisation to knots of formula (\ref{Zistorsion}) and the following lemma, whose proof makes up the remainder of this section. An alternative proof of this formula is given in \cite{DUB2}.

\begin{lemma}
Let $K$ a knot and $\rho$ an abelian representation of $\Pi_1({S}^3\setminus K)$ with conjugacy class labelled by $\theta$. Then
\begin{equation}
\mathrm{tor}\left( {S}^3\setminus \mathrm{interior} (T(K)) \right) = \frac{4 \, \sin^2\left(\tfrac{1}{2}\theta\right)}{\lvert A_K\left( e^{i\theta}\right) \rvert^2}
\end{equation}
\end{lemma}

\begin{figure}[h]
$$\epsfbox{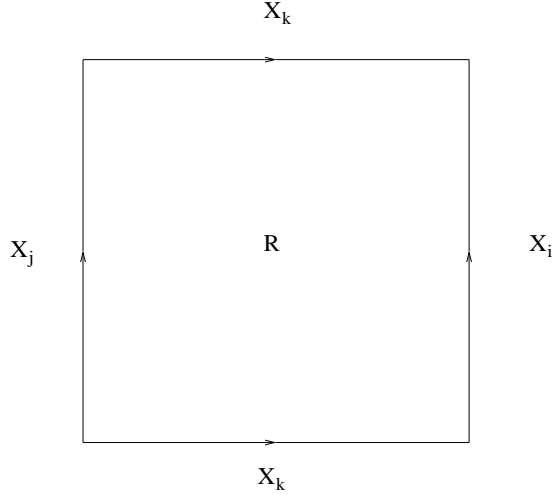}$$
 \caption{The 2-cell $R$ corresponding to the relation $x_k\,x_i\,x_k^{-1}\,x_j^{-1}$.}
 \label{typ2cell}
\end{figure}

\noindent\emph{Proof.}
This is just a generalisation of the calculations seen in section \ref{calculationoftor}. With the notation introduced in section \ref{Reidemeistertorsion}, let $\rho$ be the abelian representation of $\Pi_1({S}^3\setminus K)$ with 
\begin{equation*}
\rho(x_i) = x, \quad i = 1,2,3, \quad\quad
x = \begin{pmatrix}
	e^{i\theta/2} & 0\\
	0 & e^{-i\theta/2}
	\end{pmatrix}.
\end{equation*}	
Orient the 1-cells so that their distinguished vertices are at their `from' ends. 
A typical relation is of the form $x_k\,x_i\,x_k^{-1}\,x_j^{-1}$. The corresponding 2-cell is shown in figure \ref{typ2cell} and is attached as indicated by the labels on the 1-cells in its boundary. Choose the bottom left hand vertex to be the distinguished vertex and orient the 2-cell so that its boundary is oriented in an anti-clockwise direction.
Choose bases 
\begin{align*}
&\mathbf{c}^0 = \lbrace \Phi_1,\Phi_2,\Phi_3 \rbrace, \; \Phi_i(V) = \sigma_i \\
&\mathbf{c}^1 = \lbrace \Psi_{11}, \Psi_{21}, \Psi_{31}, \ldots , \Psi_{1n}, \Psi_{2n}, \Psi_{3n} \rbrace, \; \Psi_{ij}(X_k) = \sigma_i \delta_{jk}\\
&\mathbf{c}^2 = \lbrace \Omega_{11},  \Omega_{21}, \Omega_{31}, \ldots , \Omega_{1,n-1}, \Omega_{2, n-1}, \Omega_{3, n-1}  \rbrace, \; \Omega_{ij}(R_k) = \sigma_i \delta_{jk}.
\end{align*} 
Then we have
\begin{equation}
\dd_0\Phi_i = (x-1) \sum_{j=1}^n \Psi_{ij}.
\end{equation}
from which we see we may always choose
\begin{equation}
\mathbf{b}^0 = \lbrace \Phi_2, \Phi_3\rbrace, \quad \tilde{\mathbf{h}}^0 = \Phi_1. 
\end{equation}

Suppose now that the 2-cell in figure \ref{typ2cell} is the one which corresponds to the $l$th relation.
For a general element $\Psi_{pq} \in \mathbf{c}^1$ we have
\begin{equation*}
\dd_1 \Psi_{pq}( R_l ) = (1-x) \, \triangleright \sigma_p \, \delta_{qk} \, + \, x \, \triangleright \sigma_p \, \delta_{qi} \, - \, \sigma_p \, \delta_{qj} \, .
\end{equation*}
From this we learn that, with respect to the bases $\mathbf c^1$ and $\mathbf c^2$, the matrix for $\dd_1$, thought of as an $(n-1)\times n$ matrix, has the following entries in its $l$th row: In the $i$th column, $X$ (the $\SO(3)$ element corresponding to $x$), in the $j$th column, -1, in the $k$th column, $1-X$ and in all other columns, 0. Define $M^{(2)}$ to be the $3(n-1)\times 3(n-1)$ matrix formed by deleting the first three columns from the matrix for $\delta^1$. Now $M^{(2)}$ is constructed in precisely such a way that the $(n-1)\times(n-1)$ matrix formed by treating $X$ as an indeterminate has determinant equal to the Alexander polynomial of $K$ \cite{A}. So, since the matrix $X$ has eigenvalues 1, $e^{i\theta}$ and $e^{-i\theta}$, 
\begin{equation*}
\lvert M^{(2)} \rvert = \lvert A_K(1) \, A_K(e^{i\theta}) \, A_K(e^{-i\theta})\rvert  = \lvert A_K(e^{i\theta})\rvert^2\,,
\end{equation*}
which is non-zero by our original assumption. This means $M^{(2)}$ has rank $3(n-1)$ and we may choose
\begin{equation*}
\mathbf{b}^1 = \lbrace \Psi_{ij} | j \neq 1 \rbrace \,.
\end{equation*}
Since the elements of each row sum to zero, a basis of $Z^1$ is $\lbrace \sum_{j=1}^n \Psi_{ij} | i = 1, 2, 3 \rbrace$, and since $B^1$ is the span of $(x-1)( \sum_{j=1}^n \Psi_{1j}) \equiv 0$, $ (x-1)(\sum_{j=1}^n \Psi_{2j})$ and $ (x-1)(\sum_{j=1}^n \Psi_{3j})$, we may choose
\begin{equation*}
\tilde{\mathbf{h}}^1 = \sum_{j=1}^n \, \Psi_{1j} \,.
\end{equation*}
With all these ingredients we may calculate
\begin{equation*}
\tau_2 = \left[ \dd_1(\mathbf{b}^1) / \mathbf{c}^2 \right]^{-1} = \lvert M^{(2)} \rvert^{-1} = \frac{1}{\lvert A_K\left(e^{i\theta}\right) \rvert^2} \,.
\end{equation*}
Next we have
\begin{equation*}
\tau_1 = \left[ \dd_0(\mathbf{b}^0), \tilde{\mathbf{h}}^1,\mathbf{b}^1 / \mathbf{c}^1 \right] = \lvert M^{(1)} \rvert\,,
\end{equation*}
where $M^{(1)}$ is the $3n \times 3n$ matrix
\begin{equation*}
M^{(1)} = \begin{pmatrix}
	R & \hdotsfor{2} & R\\
	0 & 1 & \\
	\vdots & & \ddots &\\
	0 & & & 1
	\end{pmatrix},
\end{equation*}
So 
\begin{equation}
\tau_1 = \lvert R \rvert = 4 \sin^2\left(\tfrac{1}{2}\theta\right). 
\end{equation}
Finally
\begin{equation*}
\tau_0 = \left[ \tilde{\mathbf{h}}^0,\mathbf{b}^0 / \mathbf{c}^0 \right]^{-1} =  1.
\end{equation*}
and
\begin{equation*}
\rm{tor} = \frac{4\sin^2\left(\tfrac{1}{2}\theta\right)}{\lvert A_K\left(e^{i\theta}\right) \rvert^2}.
\end{equation*}

\section{Conclusion}\label{conclusion}

The paper has given a systematic definition of the Ponzano-Regge model, exploring regularisations of the original formulation of the model in terms of \sixj s and the inclusion of observables. However to provide finite answers in as many cases as possible, it was necessary to reformulate the model in terms of integrals over group variables. In these variables, the cohomology condition which guarantees the finiteness of the partition function can be stated.

The constructions given here have many parallels with the formulation of the corresponding functional integral by Witten \cite{W2}. The functional integral, without cosmological constant, reduces to an integral of the Ray-Singer analytic torsion over the space of flat connections. Since the analytic torsion is equal to the Reidemeister torsion, the results with the functional integral can be compared with the combinatorial results in this paper. Indeed, the criterion in \cite{W2} for the partition function to be finite is the non-existence of certain `zero-modes' of the frame field. These zero modes lie in the first twisted cohomology group; however using Poincar\'e duality (integration by parts), the non-existence of these is equivalent to our vanishing criterion for $H^2$. However the comparison between the two papers is not exact, because \cite{W2} considers closed manifolds without observables, whereas here we consider essentially the manifold with boundary where the observable graph has been removed. In our case, the smallest observable $\Gamma_\bullet$ is a single point, which means that we are never considering a closed manifold. The difference lies in the fact that $H^3$ is zero here, but not always in \cite{W2}. Therefore we actually reach the opposite conclusion about the manifolds (with $\Gamma_\bullet$ is our case) for which the partition function is well-defined. In this paper, this is for 3-manifolds where the flat connections are always abelian. It would also be nice to compare the results with the work of Carlip and Cosgrove \cite{CC}, but the explicit calculations in that paper concentrate on the cases where the cohomology condition is {\em not} satisfied and the partition function is infinite. It may be possible that there is some extra regularisation procedure for the combinatorial case which gives finite answers when $H^2\ne0$. It would be an interesting project to give a definition for this case. The comparison with limits of the Turaev-Viro partition function is also another interesting area for future study.

\section{Appendix 1}
In this section the definition of twisted cohomology given in the text is shown to be equivalent to the usual definition in terms of covering spaces. The formulation given in the text is local in character but depends on the entire connection. The usual definition requires only the more gauge-invariant holonomies of loops, but the construction is somewhat non-local, requiring the use of the covering space. 

Let $L$ be a cell complex (e.g. as described in section \ref{group}).  
Pick a basepoint $*$, one of the vertices of the cell complex. In addition, let $v(\sigma)$ be a distinguished vertex in each cell $\sigma$. If $\sigma$ is a cell of $L$, and $\gamma$ a path of dual edges (1-cells) from $*$ to $v(\sigma)$, then the pair $(\sigma,\gamma)$ determines a simplex $\widehat\sigma$ of the covering space $\widehat L$, two homotopic paths giving the same simplex $\widehat\sigma$. Thus $\widehat\sigma=(\sigma,[\gamma])$, where $[\gamma]$ is the homotopy equivalence class (fixing the endpoints) of $\gamma$.

An assignment of group elements to oriented 1-cells of $L$, as described section \ref{group}, is a connection, $\rho$. The connection is flat if the holonomy of every 2-cell is the identity in $\SU(2)$. A flat connection determines a homomorphism $\alpha$ from $\pi_1(L)$ into $\SU(2)$ by 
$$\alpha([\omega])=H(\omega).$$

Here an element of $\pi_1$ is represented by a path of dual edges $\omega$ from $*$ to $*$, and the composition $\omega_1\omega_2$ consists of concatenating the sequences of edges: if $\omega_1=(f_1,f_2,\ldots,f_N)$, $\omega_2=(f_{N+1},f_{N+2},\ldots,f_{N'})$, then $\omega_1\omega_2=(f_1,f_2,\ldots,f_{N'})$.

The standard definition of twisted cohomology uses the following chain groups. Take the (untwisted) cochain group $C^k\bigl(\widehat L\bigr)$ of functions from the set of $k$-simplexes of $\widehat L$ into the Lie algebra $\su(2)$, and then define the subset of cochains $C^k_\alpha(\widehat L)$ which are invariant under $\alpha$. This means that 
\begin{equation}\label{twisted}
\widehat\phi\bigl((\sigma,[\omega\gamma])\bigr)=\alpha([\omega])\action\widehat\phi(\sigma,[\gamma]).
\end{equation}

\begin{definition} The twisted cohomology of $L$ and $\rho$ is the homology of the chain groups $C^k_\alpha(\widehat L)$ with the standard coboundary operator for simplicial homology on $\widehat L$.
\end{definition}

We establish the equivalence with the definition given in section \ref{twistedcohomology} by establishing isomorphisms 
$$C^k(L)\to C^k_\alpha(\widehat L),$$
for each $k$, which commute with the respective boundary operators.

The isomorphisms are given by
$$\phi\mapsto \widehat\phi,$$
where $\widehat\phi$ is defined by
\begin{equation}\label{isomorphism}
\widehat\phi\bigl((\sigma,[\gamma])\bigr)=H(\gamma)\action\phi(\sigma).
\end{equation}
This obeys relation (\ref{twisted}). The coboundaries are given by (\ref{coboundary}) in the case of $\phi$,
\begin{equation*}
\dd \phi(\sigma)= \sum h_{v(\sigma),v(\tau)} \action\phi(\tau),
\end{equation*}
where $\partial\sigma=\sum\tau$, and in the case of $\widehat\phi$,
$$\dd \widehat\phi\bigl((\sigma,[\gamma])\bigr)=\widehat\phi\bigl(\sum(\tau,[\gamma\gamma_{v(\sigma),v(\tau)}])\bigr)
=H(\gamma)\action\sum h_{v(\sigma),v(\tau)} \action\phi(\tau).$$
These two expressions are related by the isomorphism (\ref{isomorphism}) of chain groups, and so the chain complexes are isomorphic. This induces isomorphisms on the cohomology.

\section{Appendix 2}

The decomposition of a space as a CW complex can be changed without changing the Reidemeister torsion. In fact the changes on the 2-complex $K$ can be described by a sequence of moves on the presentation of $\pi_1$ it determines. The moves on a presentation which do not change the simple homotopy type, and hence do not change the Reidemeister torsion, are known as $Q^{**}$ transformations \cite{cxbook}. Therefore to show that a 2-complex can be used to calculate the Reidemeister torsion, it is sufficient to exhibit a sequence of these $Q^{**}$ transformations.

\begin{lemma}
There exists a sequence of $Q^{**}$ transformations between the Wirtinger and `region' presentations of $\Pi_1({S}^3 \setminus K)$.
\end{lemma}

\noindent\emph{Proof.}
This makes use of an intermediate presentation called the `edge' presentation. It is in two parts; the first gives a sequence of $Q^{**}$ transformations between the `edge' and `region' presentations, the second between the `edge' and Wirtinger presentations. 

For the unknot diagram with no crossings, the two presentations are the same. In the following, this case is excluded, so it is assumed that there is at least one crossing in the knot diagram.

The edge presentation for $\Pi_1({S}^3 \setminus K)$ is defined as follows. Consider an oriented knot diagram for $K$. This diagram determines a graph $\Gamma(K)$, called the knot shadow, by replacing each crossing by a vertex. As the knot has at least one crossing, the edges of this graph divide the knot into segments, which will also be referred to as edges.

\begin{figure}[h]
$$\epsfbox{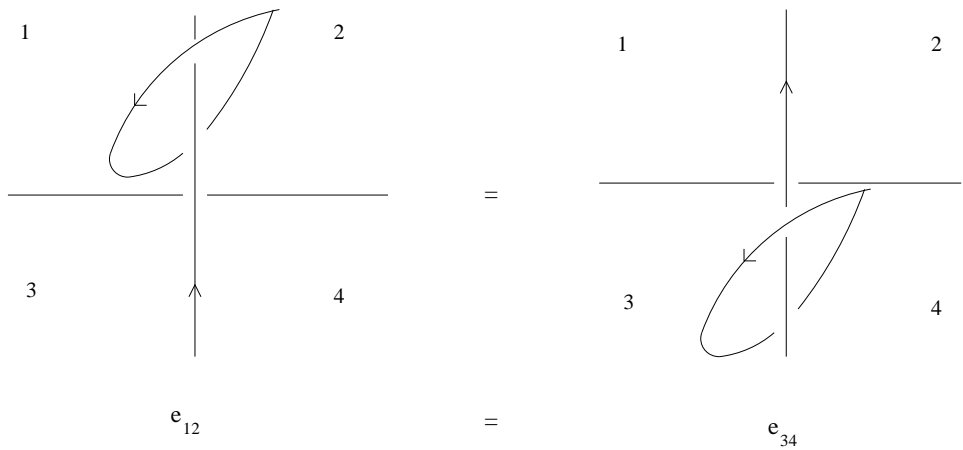}$$
 \caption{$R^{(1)} = \epsilon_{12}\epsilon_{34}^{-1}$}
 \label{type1relation}
\end{figure}

\begin{figure}[h]
$$\epsfbox{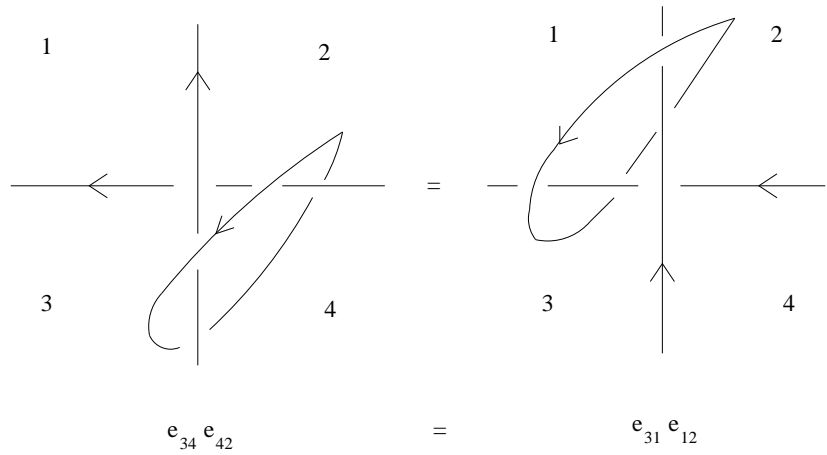}$$
 \caption{$R^{(2)} = \epsilon_{31}\epsilon_{12}\epsilon_{42}^{-1}\epsilon_{34}^{-1}$}
 \label{type2relation}
\end{figure}

Then for each edge $e_{ij}$ of the knot, bounded by the regions $i$ and $j$ of the diagram on its left and right respectively, there is a generator $\epsilon_{ij}$ of $\Pi_1({S}^3 \setminus K)$ determined by the loop encircling $e_{ij}$ in the direction of a left-handed screw. Write $\epsilon_{ji}$ for $\epsilon_{ij}^{-1}$.  At each crossing, two relations hold. The `type 1' relation is shown in figure \ref{type1relation} and the `type 2' relation in figure \ref{type2relation}.

The type 2 relations are not independent; one of the relations can always be written in terms of the others.

The relations for the edge presentation are all of the type 1 relations and all but one of the type 2 relations (it doesn't matter which we exclude). For later reference, the crossing for which the relation is omitted is labelled $X$.

Now we are ready for the proof. For the first part we start with the `edge' presentation and give a sequence of $Q^{**}$ transformations to the `region' presentation (defined in section \ref{calculationofZ}). The first step is to introduce the region generators. For this we need to choose how to define them in terms of edge generators.

\begin{figure}[h]
$$\epsfbox{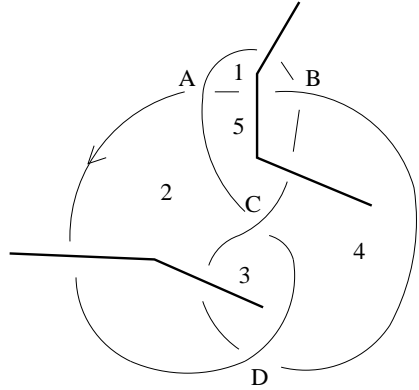}$$
 \caption{A possible choice of trees $\mathcal{T}$ for the figure-eight knot}
 \label{figeighttrees}
\end{figure}

The shadow of the knot diagram has a dual graph $\Gamma^{*} (K)$, the dual to $\Gamma(K)$ in the knot diagram. This has a vertex in each region and a dual edge $e_{ij}^*$ corresponding to each edge 
$e_{ij}$ of $\Gamma(K)$.
From the dual graph choose a subset of bivalent trees $\mathcal{T}$ which taken together visit every region of the knot diagram exactly once and each of which visits the external region exactly once. A possible choice of trees for the figure-eight knot is shown in figure \ref{figeighttrees}. 

Then, if a tree passes through the regions $0, i_1, i_2, \ldots, i_n$ in that order, the region generators $\rho_{i_1}, \rho_{i_2}, \ldots, \rho_{i_n}$ will be defined as
\begin{equation*}
\rho_0=1, \rho_{i_1} = \epsilon_{i_10}, \quad \rho_{i_2} = \epsilon_{i_2i_1}\epsilon_{i_10}, \quad\ldots \quad \rho_{i_n} = \epsilon_{i_ni_{n-1}}\ldots \epsilon_{i_10}
\end{equation*}
The introduction of the region generators defined in this way is implemented by the following $Q^{**}$ transformations. For each region $i_k$ introduce a generator $\rho'_{i_k}$ and a `type 3' relation $R^{(3)}_{i_k} = \rho'_{i_k}$. Now replace $\rho'_{i_k}$ by $\rho_{i_k}$ defined by
\begin{equation*}
\rho'_{i_k} = \epsilon_{i_ki_{k-1}}\ldots \epsilon_{i_10}\, \rho_{i_k}^{-1}
\end{equation*}
Next we want to turn the type 1 relations into the relations of the region presentation. That is, we want to write each generator $\epsilon_{jk}$ in the type 1 relations as $\rho_j\rho_k^{-1}$, with $\rho_0=1$.  There are two cases. 
\begin{itemize}
\item If  $e_{jk}^*$ belongs to $\mathcal{T}$, the set of trees defining the $r_i$, we proceed as follows. Of the two type 3 relations defining $\rho_j$ and $\rho_k$, suppose that $R^{(3)}_j$ contains more generators. Then we can use $R^{(3)}_j$ to express $\epsilon_{jk}$ as $\rho_j \, \epsilon_{0i_1}\cdots\epsilon_{i_nk}$ for some edge generators $\epsilon_{0i_1}\cdots\epsilon_{i_nk}$. And then we can use $R^{(3)}_k$ to express those edge generators as $\rho_k^{-1}$. If it is $R^{(3)}_k$ that has more generators we use $R^{(3)}_k$ first so that the intermediate step is $\epsilon_{jl_1}\cdots\epsilon_{l_n0}\rho_k^{-1}$. We illustrate the procedure for the figure-eight knot shown in figure \ref{figeighttrees}. The type 1 relation at $B$ is $\epsilon_{40}\epsilon_{51}^{-1}$. Post-multiply by $R^{(3)}_5 = \epsilon_{51} \epsilon_{01}^{-1} \rho_5^{-1}$ to get $\epsilon_{40}\epsilon_{01}^{-1} \rho_5^{-1}$. Conjugate by $\epsilon_{40}^{-1}$ to get $\epsilon_{01}^{-1} \rho_5^{-1}\epsilon_{40}$. Pre-multiply by the inverse of $R^{(3)}_1$ (that is, replace $R^{(3)}_1$ by its inverse, do the pre-multiplication, and then return the type 3 relation to $R^{(3)}_1$). Finally, conjugate by $\epsilon_{40}$ to get $\epsilon_{40}\rho_1 \rho_5^{-1}$. (The dual to $e_{40}$ does not belong to the trees so we cannot re-express $\epsilon_{40}$ using this method.)
\item If  $e_{jk}^*$ is \emph{not} in $\mathcal{T}$ then we need to express $\epsilon_{jk}$ as
\begin{equation*}
\epsilon_{jj_1}\epsilon_{j_1j_2}\cdots\epsilon_{j_m0}\epsilon_{0k_1}\epsilon_{k_1k_2}\cdots\epsilon_{k_nk}
\end{equation*}
where 
\begin{equation*}
\epsilon_{jj_1}\epsilon_{j_1j_2}\cdots\epsilon_{j_m0}\rho_j^{-1} \quad \text{and} \quad \epsilon_{0k_1}\epsilon_{k_1k_2}\cdots\epsilon_{k_nk}\rho_k^{-1}
\end{equation*}
are the relations defining $\rho_j$ and $\rho_k$ respectively, before we can use $R^{(3)}_j$ and $R^{(3)}_k$. The subgraph $\gamma$ formed by the edges 
\begin{equation*}
e_{0j_1}^*, \ldots, e_{j_mj}^*, e_{jk}^*, e_{kk_n}^*, \ldots, e_{k_10}^* \quad \in \quad \Gamma^{*} (K)
\end{equation*}
divides the knot diagram in two. By using the type 2 relations on the side of $\gamma$ \emph{not} containing $X$ (the crossing point whose type 2 relation does not appear in the presentation), we can achieve our desired re-expression of $\epsilon_{jk}$. Again, this is best understood by means of an example and we refer to the figure-eight knot pictured in figure \ref{figeighttrees}, with the excluded relation being $X = D$. As the example, choose the type 1 relation at $D$, which is $R^{(1)}_D = \epsilon_{34}\epsilon_{20}^{-1}$. The dual to $e_{34}$ does not belong to $\mathcal{T}$. For this edge, the graph $\gamma$ is defined by the edges $e_{02}^*, e_{23}^*, e_{34}^*, e_{45}^*, e_{51}^*, e_{10}^*$. Using the type 2 relation at $C$, $R^{(2)}_C = \epsilon_{32}\epsilon_{25}\epsilon_{45}^{-1}\epsilon_{34}^{-1}$, we can express $\epsilon_{34}$ as $\epsilon_{32}\epsilon_{25}\epsilon_{45}^{-1}$. Simply pre-multiply $R^{(1)}_D$ by $R^{(2)}_C$. Then using $R^{(2)}_A$ we can express $R^{(1)}_D$ as $\epsilon_{32}\epsilon_{20}\epsilon_{01}\epsilon_{51}^{-1}\epsilon_{45}^{-1}\epsilon_{20}^{-1}$. The required $Q^{**}$ transformations are easy to work out. Next replace $R^{(3)}_3$ by its inverse and use it to pre-multiply $R^{(1)}_D$ which will now read $\rho_3\epsilon_{01}\epsilon_{51}^{-1}\epsilon_{45}^{-1}\epsilon_{20}^{-1}$. Conjugate by $\epsilon_{20}^{-1}$, post-multiply by $R^{(3)}_4$ and finally conjugate by $\epsilon_{20}$ to obtain $\rho_3\rho_4^{-1}\epsilon_{20}^{-1}$. 
\end{itemize}

The final step is to eliminate the edge generators. The $\epsilon_{jk}$ for which  $e^*_{jk}$ belongs to $\mathcal{T}$ will be eliminated using the type 3 relations, the rest using the type 2 relations. 

Let $S$ be the set of edges of $\Gamma(K)$ whose duals do not belong to members of $\mathcal{T}$. It is easy to see that $S$ is a connected tree which meets every crossing. This follows because $S$ is obtained from the planar knot diagram by removing region 0 and then collapsing the diagram along the trees $\mathcal{T}$ - removing the regions and edges dual to the trees. Collapsing preserves connectedness and the Euler number of one, which means the graph $S$ is a tree. 

Pick an edge $e$ incident at $X$. There is one other crossing point at which $e$ is incident; call this $Y$. The type 2 relation at $Y$ is conjugate to $\epsilon\,\epsilon_1\,\epsilon_2\,\epsilon_3$, and this is the only relation involving the generator $\epsilon$ corresponding to $e$. Therefore $\epsilon$ and this relation can be removed from the presentation using $Q^{**}$ transformations. This process can be repeated on the smaller tree(s) $S'\subset S$ obtained by removing $e$, till all edge variables in $S$ have been removed along with all type 2 relations.

For each edge whose dual does belong to $\mathcal{T}$, consider the corresponding generator as being defined by the type 3 relation with the least number of generators to which it belongs. So in the example of the figure-eight knot, the generators and the type 3 relations which we take to define them are
\begin{align*}
&\epsilon_{01} &\text{defined by} &\quad R^{(3)}_1 = \epsilon_{10}\rho_1^{-1}\\
&\epsilon_{20} &&\quad R^{(3)}_2 = \epsilon_{20}\rho_2^{-1}\\
&\epsilon_{32} &&\quad R^{(3)}_3 = \epsilon_{32}\epsilon_{20}\rho_3^{-1}\\
&\epsilon_{45} &&\quad R^{(3)}_4 = \epsilon_{45}\epsilon_{51}\epsilon_{10}\rho_4^{-1}\\
&\epsilon_{51} \: &&\quad R^{(3)}_5 = \epsilon_{51}\epsilon_{10}\rho_5^{-1}
\end{align*}
Pick a relation with the greatest number of generators. Suppose it reads $\epsilon_1\cdots \epsilon_m \, \rho_n^{-1}$. Then since $\epsilon_1$ appears only once in this relation and not in any other relations, 
the generator $\epsilon_1$ and this relation can be eliminated by $Q^{**}$ transformations.
 Do the same for all other relations with $m+1$ generators, then for all relations with $m$ generators and so on till all the edge generators have been eliminated.
This completes the first part of the proof.

For the second part we start with the edge presentation and give a sequence of $Q^{**}$ transformations to the Wirtinger presentation. Since the latter has a single generator for each arc, we must eliminate all but one of the edge generators belonging to each given arc. We do this using the type 1 relations. Pick a vertex. Suppose the type 1 relation there reads $\epsilon \, \epsilon'$. Use this relation to eliminate $\epsilon'$ from all other type 1 and type 2 relations. (The $Q^{**}$ transformations for this step are obvious). Replace $\epsilon'$ by $\epsilon'' = \epsilon \, \epsilon'$. Now eliminate $\epsilon''$ and the relation. Repeat this process for each vertex in turn. The resulting presentation is the Wirtinger presentation. 

\subsubsection*{Acknowledgement} Thanks are due to the support of the Erwin Schr\"odinger Insititute, Vienna, and to a Short Visit Grant from the QG research networking programme of the European Science Foundation.

\end{document}